\documentclass[twocolumn,10pt,journal]{IEEEtran}

\usepackage{amsmath,amssymb,amsfonts}
\usepackage[usenames, dvipsnames]{color}
\usepackage{graphicx}
\usepackage{cite}
\usepackage{mathrsfs}
\usepackage{amsthm}
\usepackage{mathtools}

\def\diag{{\,\mathrm{diag}}}

\def\rate{\mathsf{R}}

\newtheorem{prop}{Proposition}
\newtheorem{corr}[prop]{Corollary}
\newtheorem{thm}[prop]{Theorem}

\newcounter{excounter}

\newcommand{\asmrev}[1]{\textcolor{black}{#1}}
\newcommand{\rev}[1]{\textcolor{black}{#1}}
\newcommand{\revv}[1]{\textcolor{black}{#1}}
\newcommand{\revvv}[1]{\textcolor{black}{#1}}

\begin{document}

\title{Kelly Bets and Single-Letter Codes: Optimal Information Processing in Natural Systems}

\author{Alexander~S.~Moffett and Andrew W. Eckford,~\IEEEmembership{Senior~Member,~IEEE}%
\thanks{Submitted to {\sc IEEE Transactions on Molecular, Biological, and Multi-Scale Communications,} May 31, 2024. Revised: October 10, 2024, February 10, 2025, and May 13, 2025. Accepted June 9, 2025. \revvv{Material in this paper has been submitted in part for possible publication at the 2025 IEEE Global Communications Conference (Globecom), Taipei, Taiwan.}}%
\thanks{Alexander S. Moffett is with the Center for Theoretical Biological Physics, Northeastern University, Boston, MA, USA. Andrew W. Eckford is with the Department of Electrical Engineering and Computer Science, York University, Toronto, ON, Canada. Email: aeckford@yorku.ca}%
\thanks{This work was supported in part by DARPA RadioBio grant HR001117C0125, and by Discovery grant RGPIN-2016-05288 from the Natural Sciences and Engineering Research Council.}%
}

\markboth{IEEE Transactions on Molecular, Biological, and Multi-Scale Communications, To Appear}%
{Moffett and Eckford: Kelly Bets and Single Letter Codes}

\maketitle

\begin{abstract}
In an information-processing investment game, such as the growth of a population of organisms 
in a changing environment, Kelly betting maximizes the expected log rate of growth. In this paper, we show that Kelly bets are closely related to optimal single-letter codes \revv{(i.e., they can achieve the rate-distortion bound with equality)}. Thus, natural information processing systems with limited computational resources can achieve information-theoretically optimal performance. We show that the rate-distortion tradeoff for an investment game has a simple linear bound, and that the bound is achievable at the point where the corresponding single-letter code is optimal. This interpretation has two interesting consequences. First, we show that increasing the organism's portfolio of potential strategies can lead to optimal performance over a continuous range of channels, even if the strategy portfolio is fixed. Second, we show that increasing an organism's number of phenotypes \rev{(i.e., its number of possible behaviours in response to the environment) can lead to higher growth \asmrev{rate}, and we give conditions under which this occurs. Examples illustrating the results in simplified biological scenarios are presented.}
\end{abstract}

\section{Introduction}

\IEEEPARstart{I}{t} is a commonly held view that information-theoretic performance limits are difficult to achieve, in the sense that they require complicated encoding and decoding schemes. Taking this viewpoint, information-theoretic results seem relevant only to carefully designed systems with ample computational resources, and apply only loosely to naturally occurring systems, particularly biological systems. However, a remarkable result by Gastpar and others \cite{gastpar2003code} showed that under some conditions, \revv{it is possible to achieve the rate-distortion bound with equality} using a {\em single-letter code}, \revv{and in this sense, the communication system is optimal}. As a single-letter code often means a trivial code, this implies that optimality may be achieved with practically no computation. 


In this paper our focus is on a type of coding-free communication problem known as Kelly betting \cite{kelly1956new}.
Consider an {\em investment game} with one or more possible outcomes (such as the winner of a horse race), where players can invest their wealth in one or more actions (such as betting on horses). In such a game, a reward is paid proportional to the original investment, based on the value of the chosen action, paired with the actual outcome (double your money for betting on the winning horse, or zero for betting on the wrong one). \rev{In investment games such as this one, Kelly betting has three key features. First, it is {\em geometric}: rewards are proportional to investment and are compounded in subsequent rounds. Second, it involves {\em bet hedging}: since the outcome is not perfectly known to the investor, the wealth is invested in multiple outcomes according to an investment strategy; moreover, a bet hedging distribution represents a strategy that may be optimized to increase the geometric growth rate. Third, it involves {\em communication}:} investors can receive information that changes their knowledge about the probability of each outcome, and the Kelly bet is the strategy that maximizes the expected log growth rate of the player's wealth. This simplified model can be used to represent many social or natural information-processing systems, such as gambling, describing evolutionary processes, 
and investing in the stock market \cite{thorp2008kelly,cover1984algorithm,bergstrom2004shannon}. 

\rev{Importantly, the key features of Kelly betting are widely observed in biological systems. First, an organism implements its evolutionary strategy by expressing a set of behaviours known as a {\em phenotype} (equivalent to the action in an investment game). The phenotype is selected from a set of available phenotypes, which governs its ability to survive and reproduce under given conditions. \asmrev{By phenotype, we mean any plastic attribute of an organism, whether controlled by neurological, physiological, or epigenetic processes.} The capability 
of a phenotype to help a population of organisms
establish itself and withstand competition is related to its geometric growth rate, i.e. the \asmrev{time-averaged} logarithm of its population \asmrev{size}, a measure known as {\em Malthusian fitness} \cite{wu2013interpretations}. 
Second, it is widely known that genetic regulatory systems are stochastic, and cause the organism to express a distribution of phenotypes in response to identical stimuli \cite{raser2005noise}, a phenomenon known as {\em gene expression noise} \cite{munsky2012using}. In practice, this noise allows a population
to implement bet hedging through a random distribution of phenotypes \cite{yoshimura1996evolution}.
Specific examples of bet hedging are commonly observed in bacteria \cite{de2023effective}, but have also been observed in systems as diverse as diapause (delayed development in unfavourable environments) in insects \cite{menu2002bet}, extra-pair mating in birds \cite{yasui2018bet}, and seed germination in plants \cite{abley2024bet}. 
Third, and perhaps most crucially, recent work has shown that organisms receive and process information to adapt their bet-hedging strategy. For example, bacteria are known to respond to cues to adapt their bet-hedging behaviour, either cues in the environment itself \cite{mitchell2009adaptive}, or explicitly communicated by their neighbours, such as in bacterial quorum sensing \cite{miller2001quorum,moffett2022cheater}, which can modulate bet-hedging strategies \cite{striednig2022bacterial}. \asmrev{Communication can also occur through time: a bacteria can use cues drawn from its own history and stored through epigenetic means for future use \cite{veening2008bet,felsenfeld2014brief,veening2008bistability,gianella2021ecological}.} }

\rev{The Kelly-like features of geometric growth, bet-hedging, and the transmission of information, make these natural information processing systems important targets for information theory.
From a theoretical perspective, biologists have adapted Kelly's horse racing model to define the ``fitness value of information'', which formally connects the sensory information a population collects to the population growth rate \cite{donaldson2010fitness,donaldson2008phenotypic}. Further work in this direction has extended Kelly betting to better account for the complexities of biological systems \cite{tal2020adaptive}, with a focus on how a fixed sensory channel contributes to fitness. In many of these works, an explicit connection is drawn to rate-distortion theory \cite{berger1971}, in which the distortion function is adapted such that minimizing distortion is the same as maximizing fitness, and in which distortion is explicitly related to the reliability of information available to the organism \revv{(see \cite{rivoire2011value,xue2019environment,moffett2022minimal})}. Others have explored how sensory channels could be tuned to the statistics of a fluctuating environment and with a distortion function defined by fitness \cite{taylor2007information,moffett2020fitness}.}

\rev{Information-theoretic analysis of communication systems provide mathematical limits on their performance. Biological communication systems should optimize themselves through evolution; if information processing can be done more efficiently, it would confer a fitness advantage that would be transmissible \cite{endler1993some}. 
Thus, rate-distortion limits can be used to predict features of biological systems: either predicting their operating points (at the optimum), or predicting how they dynamically evolve towards the optimum \cite{bialek2012biophysics}. The ultimate goal is to use information theory as an analytical tool for making predictions about biological systems, towards which the results in this paper take an important step.}
\rev{In this context, the importance of the single-letter code is in the achievability of the information-theoretic results. If computational constraints left a large gap to information-theoretic performance limits, then the applicability of rate-distortion theory would be limited. However, Gastpar's result indicates that this need not be the case.}

While the potential importance of Gastpar's result has been discussed in the natural communication literature \cite{rivoire2011value,varshney2007optimal,gohari2016information,tishby2010information}, our objective in this paper is to show that single-letter codes are closely related to Kelly betting, and that the relationship has interesting implications, both analytical and biological. 
\rev{To be specific, we address the following open questions in this paper:
\begin{itemize}
    \item {\em When is Kelly betting optimal from a source-channel rate-distortion perspective?}
    We address this question in Theorem \ref{lem:RD-inequality} and its corollary, Corollary \ref{cor:proportional-betting}, by 
    giving a simple linear bound on the rate-distortion function $\rate(D)$, satisfied by all investment games in which log growth rate is the figure of merit. This novel approach indicates the best possible performance of any investment game, regardless of strategy. We then give conditions for $\rate(D)$ to satisfy the bound with equality; at equality, we show that the solution is both a Kelly bet, and implements an optimal single-letter code, i.e., the solution achieves optimal information processing. Placing the solution in this information-theoretic framework shows that {\em no system can have better performance,} whether it uses single-letter coding or not. Moreover, this bound is used as the foundation for our subsequent results.
    \item {\em Is optimal information processing robust to environmental changes?} An organism's information processing involves selecting an action in response to side information, through side-information-dependent activation of genes to express a particular phenotype. In Proposition \ref{prop:bijection} and the ensuing discussion, we show that as the cardinality of the side information alphabet increases, the range of rates (i.e., channel conditions) in which our bound can be achieved with equality also increases. Viewing the channel conditions as a changing environmental variable, this implies that Kelly betting and optimal information processing can be made robust to those changes.
    \item {\em Is optimal information processing adaptive?} It is well known that mutations can modify genes, which result in novel phenotypes \cite{kaessmann2010origins}. In Theorem \ref{prop:generalization} and its corollary, Corollary \ref{corr:AddedRow}, we show that the number of available phenotypes strongly impact optimal information processing: adding a new phenotype can decrease the bound on $\rate(D)$, thus making the same growth-rate performance achievable at lower rates. We give formal conditions under which this relation holds; that is, we give conditions in which a new phenotype allows an organism to better adapt to its environment.
\end{itemize}
Our answers to these open questions can lead to novel predictions of biological behaviour. For example, our bound quantifies the organism's potential gain from optimizing its information-processing methodology, so in a stable 
environment, we would predict that {\em most organisms would operate close to our bound.} Moreover, we present simulation results indicating that operating close to the bound is feasible for many distributions of side information. 
}



The remainder of the paper is organized as follows. In Section II, we introduce investment games and Kelly betting. In Section III, we introduce rate-distortion theory, and state investment games and Kelly betting in those terms. In Section IV, we give our main results, showing which investment games have optimal single-letter codes, and describing the properties of these solutions. In Section V, we give several examples to illustrate our result, including biologically relevant examples. In Section VI, we discuss the implications and applications of our result. \revvv{In Section VII, we summarize and conclude the paper.}




\section{Kelly betting and proportional bets}

\subsection{Notation, \revv{definitions, and useful concepts}}
\label{sec:notation}

Throughout this paper, $\log$ is the natural logarithm and information is measured in nats. 

At times it will be convenient to represent probabilities as vectors or matrices. Let $P_{x,y}$ represent a $|\mathcal{Y}| \times |\mathcal{X}|$ joint probability matrix with $(i,j)$th element $\Pr(x=j, y=i)$, and similarly let $P_{x|y}$ represent a $|\mathcal{Y}| \times |\mathcal{X}|$ conditional probability matrix with $(i,j)$th element $\Pr(x=j | y=i)$. Letting $p_y = [\Pr(y=1),\ldots,\Pr(y=|\mathcal{Y}|)]$, we have that $p_x = p_y P_{x|y}$ and $P_{x,y} = \diag(p_y) P_{x|y}$. Similar definitions and expressions can be written for $P_{y|x}$. Marginal distributions can also be calculated as $p_x = \vec{1}_{|\mathcal{Y}|} P_{x,y}$ and $p_y = P_{x,y} \vec{1}_{|\mathcal{X}|}$ (where $\vec{1}_n$ is an $n$-dimensional all-one vector, which can be a row or column vector depending on context). We will also use conventional probability mass function notation, such as $p(x,y)$ for a joint probability or $p(x|y)$ for a conditional probability, to refer to individual elements of a probability mass function.

\revv{For a matrix $M$, define the null (column) space as the set of column vectors $v$ satisfying $v^T M = 0$, and the null (row) space as the set of row vectors $u$ satisfying $M u^T = 0$. The matrix $M$ is said to be a row stochastic matrix if $M$ has nonnegative entries and every row sums to 1 (the latter condition can be written as $M\vec{1} = \vec{1}$ for all-one vectors $\vec{1}$ of the appropriate size).}

\revv{Let $M^+$ be the Moore-Penrose pseudoinverse of $M$ \cite{golub2013matrix}, the unique matrix satisfying the four Moore-Penrose conditions: $M M^+ M = M$, $M^+ M M^+ = M$, $(M M^+)^T = M M^+$, and $(M^+ M)^T = M^+ M$ (specializing to the case of real-valued matrices). If $M$ has full row (resp. column) rank, then $M^+ M = I$ (resp. $M M^+ = I)$ for an identity matrix $I$ of the appropriate size, which implies that $M^+ = M^{-1}$ if $M$ is both square and full rank.}

\subsection{Investment games}

Consider an investment game where a player invests their wealth in one or more actions, and receives a reward proportional to the investment. The reward is affected by the actual state of the environment (or {\em outcome}), as well as any side information the player acquires to guide their strategy. Let $x \in \mathcal{X}$ be an outcome, let $z\in\mathcal{Z}$ be the player's action, and let $y \in \mathcal{Y}$ be side information about $x$ given to the player. We will assume that $\mathcal{X}$, $\mathcal{Y}$, and $\mathcal{Z}$ are discrete, finite sets. In biological terms, the wealth may represent the organism's population \asmrev{size}, the outcomes may represent environments in which the organism finds itself, the actions may represent phenotypes that the organism can express, and the side information may represent information gained from sensing or inter-organism communication, which allows the organism to set its bet-hedging distribution of phenotypes. \revv{This interpretation is depicted in Figure \ref{fig:Figure1}, where the organism receives information about wet or dry environments, and expresses phenotypes to match}. However, other interpretations may be possible.

An investment game is parameterized by $(R,S,p_x)$, where $R$ is the reward matrix, $S$ is the strategy matrix, and $p_x$ is the prior probability of the outcomes, defined as follows.
\begin{itemize}
    \item The reward matrix $R = [r_{zx}]$ is a $|\mathcal{Z}| \times |\mathcal{X}|$ matrix, with rows corresponding to actions, and columns corresponding to outcomes. The element $r_{zx}$ is the \revv{multiplicative} reward for action $z$ when outcome $x$ occurs: \revv{for every unit of wealth invested in $z$ and $x$, $r_{zx}$ is returned when $z$ and $x$ occur}. A valid $R$ has two properties: $r_{zx} \geq 0$ for all $z,x$ (no negative rewards); 
    and $R$ has linearly independent rows (no action is a linear combination of other actions). The latter condition implies $|\mathcal{Z}| \leq |\mathcal{X}|$. A diagonal $R$ has $|\mathcal{Z}|=|\mathcal{X}|$ and $r_{zx} = 0$ for $z \neq x$.

    \item The strategy matrix $S = [s_z^{(y)}]$ is a $|\mathcal{Y}| \times |\mathcal{Z}|$ matrix, where $s_z^{(y)}$ corresponds to the fraction of wealth invested in action $z$ with side information $y$. The $y$th row of $S$ is $s^{(y)}$, and gives the investment in each action on observing side information $y$. A valid $S$ has two properties: $s_z^{(y)} \geq 0$ (no short selling), and $\sum_z s_z^{(y)} = 1$ for each $y$ (the entire wealth is invested). Note that a valid $S$ is also a valid conditional probability given $y$.

    \item The prior probability $p_x = [\Pr(x=1),\ldots,\Pr(x=|\mathcal{X}|)]$ is a $1 \times |\mathcal{X}|$ vector of the prior probabilities of each outcome.
\end{itemize}
\rev{The outcome $x$ and side information $y$ are related through a conditional probability $P_{y|x}$, \revv{one possible interpretation of which is a communication channel through which the confederate sends information to the player, and} which is important in the rate-distortion context; see Section \ref{sec:KellyDistortion}.}

The reward $W = [w_x^{(y)}]$, per unit wealth, is given by
\begin{align}
    \label{eqn:winnings}
    W = S R .
\end{align}
%
The quantity $w_x^{(y)}$ is the multiplicative reward per unit wealth, \rev{with outcome $x$ and side information $y$}: for each dollar of the original fortune, $w_x^{(y)}$ is returned, so $w_x^{(y)}$ is the {\em rate of growth} of the wealth. The log rate of growth with outcome $x$ and side information $y$, written  $\Lambda(x,y)$, is
\begin{align}
    \label{eqn:LambdaXY}
    \Lambda(x,y) &= \log w_x^{(y)} .
\end{align}
The expected log rate of growth is given by
\begin{align}
    \label{eqn:logGrowthRate}
    \Lambda &= E\Big[\Lambda(x,y)\Big] = E\Big[\log w_x^{(y)}\Big] ,
\end{align}
\rev{taken with respect to the prior distribution $p_x$ and conditional distribution $P_{y|x}$.} \asmrev{
For an ergodic system, this is equivalent to the usual definition of Malthusian fitness \cite{rivoire2011value}
\begin{align}
\Lambda=\lim_{t\rightarrow\infty}\frac{1}{t}\log{}N_{t},
\end{align}
where $N_{t}$ is the population size at time $t$.
}

\subsection{Proportional betting strategy: The Kelly bet}
\label{sec:KellyBetting}

What strategy matrix $S$ maximizes $\Lambda$? We will deal with the  case in which $R$ is diagonal, i.e., $R = \mathrm{diag}\Big([r_{00},r_{11},\ldots,r_{|\mathcal{X}||\mathcal{X}|}]\Big)$. A diagonal reward matrix may be thought of as a winner-take-all gamble where an incorrect bet pays nothing. If $R$ is nondiagonal, or even nonsquare, it can often be reduced to the diagonal case, as we show in Section \ref{sec:MainResults}. 

With diagonal $R$,  (\ref{eqn:LambdaXY}) becomes
\begin{align}
    \label{eqn:LogSeparationDiagonal}
    \Lambda(x,y) &= \log r_{xx} + \log s_x^{(y)} ,
\end{align}
and (\ref{eqn:logGrowthRate}) becomes
\begin{align}
    \label{eqn:logGrowthRateDiagonal}
    \Lambda = E\Big[ \log r_{xx} \Big] + E \Big[ \log s_x^{(y)} \Big] .
\end{align}
The first term in (\ref{eqn:logGrowthRateDiagonal}) depends only on $R$, and can be thought of as the expected log growth rate if the player knows $x$ (or alternatively, if the side information was perfect, $y=x$). Let $\Lambda^* = E\Big[ \log r_{xx} \Big]$.
Meanwhile, the second term depends only on $S$. Recall that if $S$ is a valid strategy matrix, then it has the form of a valid conditional probability of $x$ given $y$. Then
\begin{align}
    E\Big[\log s_x^{(y)} \Big] &= 
    \sum_{x,y} p(x,y) \log s_x^{(y)}\\
    &\leq \sum_{x,y} p(x,y) \log p(x|y) \\
    &= - H(X|Y) ,
\end{align}
where equality is achieved with $s_x^{(y)} = p(x|y)$. This strategy is known as {\em proportional betting} or {\em Kelly betting}, and gives a growth rate of
\begin{align}
    \label{eqn:OptimalGrowthRateExample}
    \Lambda &= \Lambda^* - H(X|Y) .
\end{align}
Thus, the Kelly betting strategy gives the maximum possible log growth rate. Furthermore, the minimum loss compared with perfect knowledge is $H(X|Y)$.

Kelly's observation was that if the investment game gives fair odds (i.e., $r_{xx} = \frac{1}{p(x)}$), then 
\begin{align}
    \label{eqn:LambdaMutualInformation}
    \Lambda &= H(X) - H(X|Y) = I(X;Y) ,
\end{align}
see \cite{kelly1956new}.
A similar derivation is described in \cite[Ch. 6]{cover-book}. 



\section{Rate-distortion formulation and single-letter codes}

\rev{Investment games and other log-growth-rate problems can be formulated in a rate-distortion framework (see \cite{moffett2022minimal,marzen2016bio,rivoire2011value}). In this section we discuss the physical significance of this formulation, and subsequently give the mathematical details.} 

\subsection{Physical significance}


Formally, the source-channel rate-distortion setting consists of a transmitter, a channel, and a receiver. In the general setting, the transmitter has noncausal knowledge of all outcomes for all time, and uses this knowledge to encode a vector $\vec{x}$. The vector is sent over a possibly noisy channel. 
The receiver observes the vector $\vec{y}$ and decodes this channel output to obtain estimates of the source for all time.

In a biological setting, $\vec{x}$ represents an environmental situation of interest to the organism. The organism observes $\vec{y}$, a possibly noisy, possibly quantized version of $\vec{x}$. The fidelity of $\vec{y}$ given $\vec{x}$ may be described using a distortion function $d(x,y)$, where $x,y$ are individual corresponding elements of the vectors $\vec{x},\vec{y}$. (To simplify our notation, we do not use a separate symbol for the encoded source symbols or channel outputs.) Importantly, the consequences of error can be embedded in the distortion function, particularly the effect on growth rate of a correct or incorrect representation. Moreover, in a natural setting, the observation $\vec{y}$ may arise from explicit communication: many natural systems share environmental states with each other, e.g. through quorum sensing \cite{miller2001quorum,moffett2022cheater} or pheromonal signalling \cite{verheggen2010alarm}. Thus, natural communication can be understood in rate-distortion terms. 


\rev{Since the system is described in rate-distortion terms, an arbitrarily capable organism could, in principle, express the source as efficiently as possible using block encoding. That is, for a given $P_{y|x}$, we could generate a set of \revv{vector} quantization points $\vec{y}$, and assign vectors of source symbols $\vec{x}$ to them, \revv{as in the proof of the rate-distortion theorem}. Over a noise-free channel, the transmitter would send $\vec{y}$, which would be more efficient than sending $\vec{x}$. An example is given in Section \ref{sec:examples}; see also \cite[Ch. 10]{cover-book}. While designs have been presented to achieve the rate-distortion bound with block quantization schemes \cite{martinian2006low}, these schemes have high computational complexity, which are likely beyond the capabilities of natural and biological systems to implement,} \revv{even if noncausal access to $\vec{x}$ was available}.

\begin{figure*}[t!]
    \centering
    \includegraphics[width=0.9\textwidth]{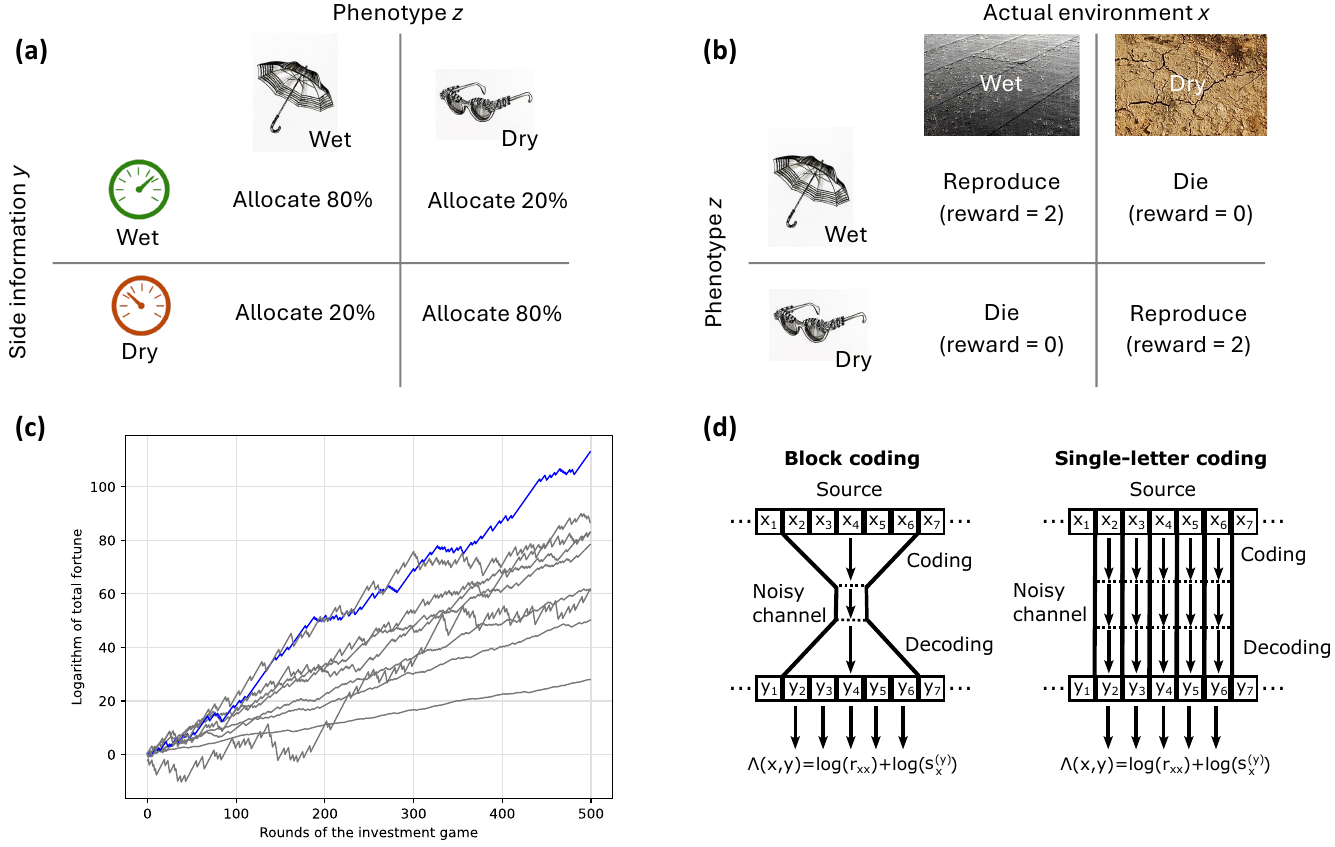}
    \caption{\revv{Depiction of the physical and communication model. {\em Subfigures (a)-(b):} The strategy matrix $S$ (subfigure (a)) and reward matrix $R$ (subfigure (b)) for an investment game, in which an organism must express a phenotype appropriate to either a wet or dry environment. The organism receives noisy side information $y$ about the environment (depicted by gauges), expresses a mixed allocation of wet and dry phenotypes (represented by an umbrella or pair of sunglasses, respectively), which then encounter the true environment and receive the appropriate reward. {\em Subfigure (c):} An example of the investment game when the side information has probability $0.8$ of being accurate. \revvv{Logarithm of total fortune is plotted versus round of the investment game} for the strategy from subfigure (a) (blue line), which is the optimal strategy; the performance of other strategies (grey lines) is also depicted. {\em Subfigure (d):} On the left, block coding is depicted, in which the vector $\vec{x}$ is encoded, passed along the possibly noisy channel, and decoded as the representation $\vec{y}$. On the right, single-letter coding is depicted, in which the encoder and decoder are trivial, and the instantaneous output $y$ is generated from the channel output arising from the instantaneous input $x$.}}
    \label{fig:Figure1}
\end{figure*}

\rev{However, there exists a low-complexity scheme that fits within the rate-distortion framework, called {\em single-letter coding}.}
In this scenario, only the current outcome $x$ is used, and the representation $y$ is the instantaneous channel output, while the encoding and decoding functions are trivial: that is, the instantaneous value of the source is directly transmitted across the channel, and the received value is directly and instantaneously used as the estimate of the source. \revv{The single-letter framework is depicted on the right in Figure \ref{fig:Figure1}(d). }
A key question is whether a natural communication system trades off performance for complexity, potentially motivating it to seek simple ``codes'' within its computational capabilities, or whether single-letter codes achieve optimal performance. As we describe below, the latter is true if the objective is to maximize expected log growth rate.


\subsection{Mathematical description: Proportional betting and rate distortion}
\label{sec:KellyDistortion}

\rev{Describing the rate-distortion formulation mathematically,} suppose for the moment that the channel is noise-free.  The transmitter forms a representation $y$ of the source symbols $x$ with rate $\rate$, where the distortion function $d(x,y)$ expresses the cost of representing $x$ with $y$.
With average distortion $D = E[d(x,y)]$, the system operates at a point $(D,\rate)$. The rate-distortion problem is to find the rate-distortion function $\rate(D)$, the minimum rate at which $D$ can be achieved. It is known that
\begin{align}
    \label{eqn:RD}
    \rate(D) &= \min_{P_{y|x} \in \mathcal{P}_D} I(X;Y),
\end{align}
where $\mathcal{P}_D$ is the set of conditional distributions $P_{y|x}$ such that $E[d(x,y)] \leq D$ \cite{cover-book}.


Again restricting our attention to a diagonal reward matrix $R$, log growth rate is given by
\begin{align}
    \label{eqn:LambdaT}
    \Lambda(x,y) &= \log w_x^{(y)} = \log s_x^{(y)}r_{xx} .
\end{align}
The objective is to maximize expected log growth rate $\Lambda = E[\Lambda(x,y)]$. As distortion is minimized in rate-distortion theory, we can set the distortion function as $d(x,y) = -\Lambda(x,y) = -\log w_x^{(y)}$, so that minimizing $E[d(x,y)]$ maximizes $\Lambda$.
Let $\Lambda^*(x) = \log r_{xx}$, then
\begin{align}
    \label{eqn:DistortionFunction}
    d(x,y) = -\log w_x^{(y)} = -\log s_x^{(y)} - \Lambda^*(x) .
\end{align} 
The player's wealth grows when $d(x,y)$ is negative, and shrinks when $d(x,y)$ is positive. The average distortion is given by
\begin{align}
    \label{eqn:AverageDistortion}
    E\Big[d(x,y)\Big] &= E\Big[ - \log s_x^{(y)} \Big] - E[\Lambda^*(x)] \\
    \label{eqn:AverageDistortion2}
    &= E\Big[ - \log s_x^{(y)} \Big] - \Lambda^* 
\end{align}
(cf. (\ref{eqn:OptimalGrowthRateExample})). The optimization in (\ref{eqn:RD}) is then used to find the minimum rate that achieves $D = E[d(x,y)]$.

It is important to note that {\em in our formulation, the strategy is constant}. That is, the strategy matrix $S$ is constant as a parameter of the investment game, and chosen independently of the distribution $p(y|x) \in \mathcal{P}_D$ that solves (\ref{eqn:RD}). Thus, there may exist zero, one, or many $D$ such that $S$ performs proportional betting with respect to $p(y|x)$. Postulating that the strategy is fixed is done for two reasons. First, it fixes the distortion function (\ref{eqn:DistortionFunction}), avoiding the problem of double optimization (cf. \cite{moffett2022minimal,still2010optimal}). \rev{Optimizing the strategy requires optimizing over the distortion function, whereas our use of rate-distortion theory requires the distortion function to be unchanging.} Second, in a social or biological setting it is possible that strategies would be fixed in advance and might not be explicitly optimized for each game. 
\rev{While the related concept of {\em phenotypic plasticity} considers how variable phenotypes may be deployed in different environments, it is known that bacteria often do not have this kind of adaptation \cite{agrawal2001phenotypic}; their phenotypic strategies can be considered  constant. Moreover,} in an evolutionary scenario, it is interesting to consider how fixed strategies might compete against each other.

\subsection{Noisy channels, channel matching, and single-letter codes}


So far we have assumed that the channel connecting the transmitter and receiver is noiseless, but suppose instead that the channel is noisy. \revv{One well-known strategy is to treat the source and channel separately, quantizing the source at rate $R_s$ (nats per source symbol), and encoding those quantized outputs with a channel code at rate $R_c$ (nats per channel use). 
Let $\rho = R_c/R_s$ (source symbols per channel use) be the concatenated code rate. For a fixed $\rho$, it follows that the minimum average distortion is $D$ satisfying $\rho = C / \rate(D)$, where $C$ is the Shannon capacity of the channel.} 
The source-channel separation theorem \cite[Thm. 7.13.1]{cover-book} (see also \cite{vembu1995source}) \revv{implies that this is also the minimum distortion for any system, not only those where source and channel are separated.}

\revv{Source-channel separation is sufficient, but not necessary, for optimality:} Berger \cite{berger2002living} observed that $\rate(D)$ could be achieved with a single-letter code if two conditions hold, a situation he called ``doubly matched''. First, the rate at which source symbols are generated is equal to the rate at which the channel can be used. This condition implies \revv{$\rho = 1$, so $\rate(D) = C$.} Second, there exists a single-letter encoder $r$ from source symbols $x$ to channel inputs, and decoder $w$ from channel outputs to representation symbols $y$, such that $\rate(D)$ is achieved. \revv{From these conditions and the source-channel separation theorem, the minimum distortion is given by $D$ satisfying $\rate(D) = C$; this is the significance of $\rate(D)$ in a single-letter code.}


Berger's observation is closely related to Gastpar's single-letter coding result \cite{gastpar2003code}. For a particular distortion function and channel input-output probability $p(y|x)$, a single-letter code operates at a point $(D,\rate)$ given by $D = E[d(x,y)]$ and $\rate = I(X;Y)$.
There exists a cost function $\rho(x)$ so that the capacity $C$ can be defined in terms of a maximum average cost $\Gamma$, i.e. $C(\Gamma) = \max_{p(x):E[\rho(x) \leq \Gamma]} I(X;Y)$.
From \cite[Thm. 6]{gastpar2003code}, a single-letter code is optimal if
and only if it operates at rate $\rate(D) = C(\Gamma)$, and the distortion function is of the form 
\begin{equation}
    \label{eqn:gastpar_distortion}
    d(x,y)=-c\log{}p(x|y)+d_{0}(x) ,
\end{equation}
where $c$ is any positive constant, and $d_0(x)$ is any function of $x$.
The use of $C(\Gamma)$, rather than the unconstrained capacity $C$, is significant: depending on the cost function, it is generally possible to select $\Gamma$ so that $C(\Gamma) = I(X;Y)$, so an optimal solution is available for (almost) any $I(X;Y)$, if $\Gamma$ is varied; we discuss the range of solutions in the discussion, section \ref{sec:single-letter-coding-versus-sensing}.
(There is an additional requirement that the cost function satisfies \cite[Lem. 3]{gastpar2003code}, but this is not important to our results and we assume a cost function is used that satisfies the condition; we also justify this assumption in section \ref{sec:single-letter-coding-versus-sensing}. We omit other details and special cases of the result in \cite{gastpar2003code} as they are not used here.) 

\section{Main Results}
\label{sec:MainResults}

\rev{In this section we give our main results. The results begin with an important preliminary result of linear algebra, Proposition \ref{lem:decomposition}, which allows the transformation of a nondiagonal or nonsquare reward matrix into a diagonal reward matrix with a modified (equivalent) strategy. Subsequent results explicitly apply rate-distortion theory to investment games, and explore the information-theoretic properties of these games.
}

\subsection{Reducing nonsquare and nondiagonal reward matrices to the diagonal case}

\rev{Consider equations (\ref{eqn:winnings})-(\ref{eqn:LogSeparationDiagonal}) when the reward matrix $R$ is nondiagonal: we still have $W = SR$, but the individual terms $w_x^{(y)}$ generally contain a summation. As a result, they cannot be easily separated by the logarithm, as in (\ref{eqn:LogSeparationDiagonal}), making it difficult to apply Kelly betting directly. However,} in this section, we show how to reduce nondiagonal and nonsquare reward matrices to the diagonal case, so that the proportional betting solution may be used. Our approach is to decompose a nondiagonal reward matrix $R$ into a product of two matrices: $R = BQ$, where $Q$ is diagonal, and $B$ is row stochastic. The diagonal matrix $Q$ is then an effective reward matrix for an equivalent investment game with hypothetical actions, while the row stochastic matrix transforms the actual strategy $S$ into an effective strategy over those actions. 

The following proposition establishes the existence and properties of the decomposition:
\begin{prop}\label{lem:decomposition}
Suppose $R$ is a valid reward matrix. Let $R^+$ represent the Moore-Penrose pseudoinverse of $R$, and let
\begin{equation}\label{eqn:decomposition-1}
q=R^+\vec{1}_{|\mathcal{Z}|} ,
\end{equation} 
where $\vec{1}_{|\mathcal{Z}|}$ is an all-one vector of length $|\mathcal{Z}|$.
If 
\begin{align}
    \label{Eqn:condition-decomposition}
    q_i>0 \:\:\forall i , 
\end{align}
then there exists a decomposition
\begin{equation}\label{eqn:decomposition0}
R=BQ ,
\end{equation}
where $B$ is a row stochastic matrix, and $Q$ is a valid $|\mathcal{X}| \times |\mathcal{X}|$ diagonal reward matrix given by
\begin{align}
    \label{eqn:diagonal-Q}
    Q = \Big(\diag(R^+ \vec{1}_{|\mathcal{Z}|}) \Big)^{-1}.
\end{align}
%
\end{prop}
\begin{proof}
We start by showing that $Q$ in (\ref{eqn:diagonal-Q}) makes $B=RQ^{-1}$ in (\ref{eqn:decomposition0}) row stochastic under the condition in (\ref{Eqn:condition-decomposition}). 

The matrix $B$ in (\ref{eqn:decomposition0}) has the same dimensions as $R$ (i.e., $|\mathcal{Z}|\times |\mathcal{X}|$), so it has \revv{rows that sum to 1} if it satisfies $B\vec{1}_{|\mathcal{X}|} = \vec{1}_{|\mathcal{Z}|}$.
The setting in (\ref{eqn:diagonal-Q}) makes the \revv{rows of $B$ sum to 1}: starting with
\begin{equation}\label{eqn:decomposition1}
B = RQ^{-1}= R\diag(R^+\vec{1}_{|\mathcal{Z}|}) ,
\end{equation}
and applying to (\ref{eqn:decomposition1}),
\begin{align}
B\vec{1}_{|\mathcal{X}|} =
RQ^{-1}\vec{1}_{|\mathcal{X}|} &=R\diag(R^+\vec{1}_{|\mathcal{Z}|})\vec{1}_{|\mathcal{X}|} \\ \label{eqn:decomposition2}
&=RR^+\vec{1}_{|\mathcal{Z}|} \\
\label{eqn:decomposition3}
&=\vec{1}_{|\mathcal{Z}|},
\end{align}
where (\ref{eqn:decomposition2}) follows since $\diag(v)\vec{1}_{|\mathcal{X}|}=v$ for any length-$|\mathcal{X}|$ vector $v$, and (\ref{eqn:decomposition3}) follows since $R$ has linearly independent rows, in which case $R R^+ = I$.
	
Next we show that $B$ does not contain any negative entries under condition (\ref{Eqn:condition-decomposition}). With $Q^{-1}=\diag(R^+ \vec{1}_{|\mathcal{Z}|})$, $Q^{-1}$ has a positive diagonal if every element $q_i$ of $q=R^+\vec{1}_{|\mathcal{Z}|}$ is positive. Moreover, since $Q^{-1}$ is diagonal, $(Q^{-1})_{ii} = 1/Q_{ii}$, so $Q$ has a positive diagonal if $Q^{-1}$ has a positive diagonal. 
Since $R$ is a valid reward matrix, its entries are nonnegative. Since $Q^{-1}$ has a positive diagonal under condition (\ref{Eqn:condition-decomposition}), then $B = RQ^{-1}$ is nonnegative as the product of two matrices in each of which every element is nonnegative.
Thus $B$ \revv{has rows that sum to 1} and contains no negative entries, so it is row stochastic; this proves the proposition.
%
\end{proof}

This decomposition has previously been given for square $R$ \cite{donaldson2008phenotypic}, and the decomposition is unique if $R$ is square (see also \cite{brualdi1966}). (Recall that $R^+ = R^{-1}$ if $R$ is square and full rank.) However, if the reward matrix is nonsquare (with $|\mathcal{Z}| < |\mathcal{X}|$), then the decomposition is not unique, and examples can be found by examining the null (row) space of $R$; a detailed discussion of this case is found in section \ref{sec:nonsquare}. (Although our assumptions exclude the case where $|\mathcal{Z}| > |\mathcal{X}|$, the decomposition cannot generally be found in this case.)

Equation (\ref{eqn:winnings}) can now be rewritten
\begin{align}
    \label{eqn:NondiagonalWinnings}
    W &= SBQ = TQ ,
\end{align}
where $T = SB$. We call $T$ the {\em equivalent strategy} with respect to the diagonal reward matrix $Q$. Note that $T$ is always a $|\mathcal{Y}|\times|\mathcal{X}|$ matrix; moreover, because $B$ is row stochastic, $T$ has the properties of a valid strategy matrix (i.e., row stochastic) as long as $S$ is valid. 


If $Q$ was the true reward matrix, then the proportional betting strategy would require a strategy matrix of $T = P_{x|y}$. However, the actual strategy is $S$, where $S B = T$. Can $S$ be chosen so that $T = P_{x|y}$?
Assuming the decomposition in Proposition \ref{lem:decomposition} exists, there are three cases:
\begin{enumerate}
    \item $R$ is square and diagonal (the trivial case): The decomposition in Proposition \ref{lem:decomposition} leads to $Q=R$ and $B=I$, while $T = I S = S$. Thus, we achieve proportional betting with $S = P_{x|y}$.
    \item $R$ is square but nondiagonal: Letting $T = P_{x|y}$, we obtain $S = T B^{-1} = P_{x|y} B^{-1}$. Proportional betting is achieved as long as $S$ is a valid strategy matrix, which is not always the case; see the example below.
    \item $R$ is nonsquare: For a particular $B$, the equation $S B = P_{x|y}$ cannot be solved for all $P_{x|y}$, as the equation is overdetermined for $S$; however, as noted in the discussion after Proposition \ref{lem:decomposition}, the decomposition leading to $B$ is not unique. The consequences of this case are discussed below, in Section \ref{sec:nonsquare}.
\end{enumerate}

To illustrate how the second case may lead to an invalid $S$, consider the following example: let
\begin{align}
    R &= 
    \left[
        \begin{array}{cc}
            2 & 1 \\
            1 & 3
        \end{array}
    \right] .
\end{align}
The decomposition in (\ref{eqn:decomposition0}) exists, with
\begin{align}
    B^{-1} = 
    \left[
        \begin{array}{rr}
            \frac{3}{2} & -\frac{1}{2} \\
            -1 & 2
        \end{array}
    \right] . 
\end{align}
For a particular $y$, suppose one row of $T$ is given by $t^{(y)} = \left[t,1-t\right]$. The corresponding row $s^{(y)}$ of $S$ is
\begin{align}
    s^{(y)} &= \Big[t,1-t\Big] B^{-1} = \left[\frac{5t-2}{2}, \frac{4-5t}{2}\right] .
\end{align}
It can be checked that $s^{(y)}$ is a valid strategy for $\frac{2}{5} \leq t \leq \frac{4}{5}$. Thus, if $p(x|y) > \frac{4}{5}$ for some $x$ and $y$, proportional betting is not possible because there is no actual strategy $s^{(y)}$ that can achieve it.




\subsection{Proportional betting and bounds on $\rate(D)$ for general investment games}

In light of Proposition \ref{lem:decomposition}, we rewrite the distortion function (\ref{eqn:DistortionFunction}): 
\begin{align}
    \label{eqn:GeneralizedDistortionFunction}
    d(x,y) = -\log w_x^{(y)} = -\log t_x^{(y)} - \Lambda^*(x) ,
\end{align} 
where $\Lambda^*(x) = \log q_{xx}$, the $x$th diagonal element of $Q$. This form of the distortion function is used throughout the remainder of the paper.

We now consider the channels $P_{y|x}$ in which the strategy $T$ performs proportional betting, and give an achievable bound on $\rate(D)$, which represents the gap to optimality for an arbitrary investment game. To solve this problem we need to find which joint distribution or distributions $P_{x,y}$, if any, correspond to a given $P_{x|y}$ and $p_x$. Let $\phi_{P_{x|y},p_x}$ represent the set of these $P_{x,y}$: 
%
\begin{align}
    \label{eqn:ProportionalBettingSet}
    \nonumber\lefteqn{\phi_{P_{x|y},p_x}} & \\
    &= \left\{ P_{x,y} : \diag(P_{x,y} \vec{1}_{|\mathcal{X}|})^{-1}P_{x,y} = P_{x|y}, \:\:\vec{1}_{|\mathcal{Y}|} P_{x,y} = p_x \right\} ,
\end{align}
where, using basic probability, the first condition in (\ref{eqn:ProportionalBettingSet}) ensures that all elements in the set have the given $P_{x|y}$, and the second condition ensures that they have the given $p_x$.
Our results in the remainder of the paper rely on characterizing this set.

First, we can show that all investment games satisfy a simple linear bound on $\rate(D)$, and this bound is achieved with equality for all distributions in the set given above:
\begin{thm} 
    \label{lem:RD-inequality}
    For a system with distortion function $d(x,y) = -\log t_x^{(y)} - \Lambda^*(x)$, where $T$ is a valid strategy, and for $D$ such that $\mathcal{P}_D \neq \emptyset$,
    \begin{align}
        \label{eqn:RD-bound}
        \rate(D) \geq H(X) - D - \Lambda^* .
    \end{align}
    Equality in (\ref{eqn:RD-bound}) is achieved if and only if $\phi_{T,p_x}$ is nonempty, and equality occurs at $D^* = -\sum_{x,y} p(x,y) \log t_x^{(y)} - \Lambda^*$, for each $P_{x,y} \in \phi_{T,p_x}$. 
\end{thm}
\begin{proof}
    Manipulating (\ref{eqn:RD}),
    \begin{align}
        \label{eqn:RD2}
        \rate(D) &= H(X) - \max_{p(y|x) \in \mathcal{P}_D} H(X|Y) .
    \end{align}
    Let $\Delta_{p(y|x)} = E_{p(y|x)}[d(x,y)]$ represent average distortion calculated with the conditional distribution $p(y|x)$ and the given prior distribution $p_x$. For every $p(y|x) \in \mathcal{P}_D$, $D \geq \Delta_{p(y|x)}$ by definition. Since $\mathcal{P}_D \neq \emptyset$ by assumption, we can calculate $\Delta_{p(y|x)}$ of an element in $\mathcal{P}_D$ as 
    \begin{align}
        \Delta_{p(y|x)} &= E_{p(y|x)}[d(x,y)] \\
        &= \sum_{x,y} p(y|x)p(x) \Big( -\log t_x^{(y)} - \Lambda^*(x) \Big) \\
        \label{eqn:Lemma2Proof-1}
        &\geq \sum_{x,y} p(y|x)p(x) \Big( -\log p(x|y) \Big) - \Lambda^* \\
        &= H(X|Y) - \Lambda^* .
    \end{align}
    In (\ref{eqn:Lemma2Proof-1}), $p(x|y)$ is the conditional distribution associated with $p(y|x)p(x)$; using this distribution, 
    the inequality follows from the properties of entropy, since $t_x^{(y)}$ forms a valid conditional distribution $p(x|y)$. Furthermore, $\Lambda^*(x)$ is only a function of $x$ and is thus invariant to the particular $p(y|x) \in \mathcal{P}_D$ that was chosen.
    Continuing the argument, $D \geq \Delta_{p(y|x)} \geq H(X|Y) - \Lambda^*$ for any element of $\mathcal{P}_D$, so it is also true of the maximal element, i.e.,
    \begin{align}
        \label{eqn:D-condition}
        D + \Lambda^* \geq \max_{p(y|x) \in \mathcal{P}_D} H(X|Y) ,
    \end{align}
    which can be substituted into (\ref{eqn:RD2}). The inequality (\ref{eqn:RD-bound}) follows.
    
    To show the equality condition, 
    first consider ``if'': suppose $\phi_{T,p_x}$ is nonempty, and let $P^*_{x,y} \in \phi_{T,p_x}$ be one of its elements. By the definition of $\phi_{T,p_x}$, the corresponding conditional distribution is $P_{x|y}^* = T$. Furthermore,
    \begin{align}
        \nonumber
        D^* + \Lambda^* &= -\sum_{x,y} p^*(x,y) \log t_x^{(y)}\\ &= -\sum_{x,y} p^*(x,y) \log p^*(x|y) = H^*(X|Y) ,
    \end{align}
    where $p^*(x,y)$ and $p^*(x|y)$ are elements chosen from $P_{x,y}^*$ and $P_{x|y}^*$, respectively; and
    where $H^*(X|Y)$ is the conditional entropy corresponding to $p^*(x|y)$.
    By the same argument as in the first part, $D^* + \Lambda^* \geq \Delta_{p(y|x)} \geq H(X|Y)$ for any element of $\mathcal{P}_{D^*}$, and $D^* + \Lambda^* = H^*(X|Y)$ for at least one element of $\mathcal{P}_{D^*}$, since $\phi_{T,p_x}$ is nonempty. Thus, $D^* + \Lambda^* = \max_{p(y|x) \in \mathcal{P}_{D^*}} H(X|Y)$, and equality is achieved in (\ref{eqn:RD-bound}).

    Now consider ``only if'': suppose $\phi_{T,p_x}$ is empty, i.e., there is no joint distribution $P_{x,y}^*$ corresponding to $T$ and $p_x$. For any $p^*(y|x)\in\mathcal{P}_{D^*}$, where $p^*(x|y)$ is the corresponding conditional distribution, this implies
    \begin{align}
        \nonumber
        H(X|Y) &= -\sum_{x,y} p^*(y|x)p(x) \log p^*(x|y)\\ 
        &< -\sum_{x,y} p^*(y|x)p(x) \log t_x^{(y)} ,
    \end{align}
    where the strict inequality follows since $t_x^{(y)} \neq p^*(x|y)$. Thus, $D^* + \Lambda^* \geq \Delta_{p^*(y|x)} > H(X|Y)$ for any element of $\mathcal{P}^*$, which also applies to the maximal element, so equality cannot be achieved in (\ref{eqn:RD-bound}).  
\end{proof}
%


%
%
%
%


The above result leads immediately to the following:
\begin{corr}
    \label{cor:proportional-betting}
    For every $P_{x,y} \in \phi_{T,p_x}$, the system implements proportional betting (i.e., Kelly betting), and the corresponding single-letter code is optimal. 
\end{corr}
\begin{proof}
    If $P_{x,y} \in \phi_{T,p_x}$, then by definition $P_{x|y} = T$, so proportional betting is performed. Moreover, the distortion function (\ref{eqn:GeneralizedDistortionFunction}) satisfies (\ref{eqn:gastpar_distortion}) with $c=1$ and $d_0(x) = \Lambda^*(x)$. Along with our assumptions that $I(X;Y) = C(\Gamma)$, and that the cost criterion is satisfied, this is sufficient to show that the corresponding single-letter code is optimal, by \cite[Thm. 6]{gastpar2003code}.
\end{proof}
%
From this corollary, the optimality of an investment game can be easily checked by observing whether it achieves the simple linear bound given in (\ref{eqn:RD-bound}); moreover, if it does not, then the gap to the bound is the penalty for not using proportional betting.



\subsection{Properties of $\phi_{T,p_x}$ and the strategy matrix $T$}
\label{sec:StrategyMatrixProperties}

From the above results, the set $\phi_{T,p_x}$ is the key to understanding when an investment game $(R,S,p_x)$ is optimal, and is associated with an optimal single-letter code. Here we present results that characterize this set and its implications for optimal strategies.

In (\ref{eqn:ProportionalBettingSet}), the set $\phi_{T,p_x}$ was constructed explicitly, ensuring every element had conditional distribution $P_{x|y} = T$ and marginal distribution $p_x$.
In the following result, we simplify the search for solutions, showing that $\phi_{T,p_x}$ can be formulated as the set of solutions to a single linear equation.

\begin{prop}
    \label{prop:bijection} \revv{Let $\phi_{T,p_x}$ be the set defined in (\ref{eqn:ProportionalBettingSet}).} Then
    $P_{x,y} \in \phi_{T,p_x}$ if and only if $P_{x,y} = \diag(p_y) T$, where $p_y$ is a solution to the equation
    \begin{align}
        \label{eqn:py-solutions}
        p_y T = p_x , 
    \end{align}
    \revv{where $P_{x,y}$, $p_x$, and $p_y$ are as defined in Section \ref{sec:notation}.}
\end{prop}
%
%
\begin{proof}
    Let
    \begin{align}
        \psi_{T,p_x} = \{ p_y : p_y T = p_x \}.
    \end{align}
    We prove the proposition by showing that the sets $\phi_{T,p_x}$ and $\psi_{T,p_x}$ are isomorphic. For $p_y \in \psi_{T,p_x}$, let $Q = \mathrm{diag}(p_y)T$; we will show that this relation forms a bijection between the sets. First $\psi_{T,p_x} \rightarrow \phi_{T,p_x}$: if $p_y \in \psi_{T,p_x}$, then
    \begin{align}
        Q \vec{1}_{|\mathcal{X}|} = \mathrm{diag}(p_y) T \vec{1}_{|\mathcal{X}|} = \mathrm{diag}(p_y) \vec{1}_{|\mathcal{Y}|} = p_y ,
    \end{align}
    where the second equality follows from the fact that every row of $T$ sums to 1 (as it is a valid strategy), and the third equality follows from the same argument as in (\ref{eqn:decomposition2}). Substituting into the first condition in (\ref{eqn:ProportionalBettingSet}), we have $\diag(p_y)^{-1}Q = T$, which is satisfied by $Q = \mathrm{diag}(p_y)T$. As for the second condition in (\ref{eqn:ProportionalBettingSet}),
    \begin{align}
        \label{eqn:bijection1}
        \vec{1}_{|\mathcal{Y}|}Q = \vec{1}_{|\mathcal{Y}|}\mathrm{diag}(p_y)T = p_y T = p_x,
    \end{align}
    where the second equality again follows from (\ref{eqn:decomposition2}), and the third equality follows by definion of $\psi_{T,p_x}$. Thus, we have a mapping $\psi_{T,p_x} \rightarrow \phi_{T,p_x}$. Now $\phi_{T,p_x} \rightarrow \psi_{T,p_x}$: given $Q$ and a valid $T$, any $Q \in \psi_{T,p_x}$ can be written $Q = \mathrm{diag}(p_y)T$, by elementary probability, for some $p_y$; from (\ref{eqn:ProportionalBettingSet}) and (\ref{eqn:bijection1}), $p_y \in \psi_{T,p_x}$. Thus, the given relation is a bijection, and the proposition follows.
\end{proof}

Considering the possible solutions of (\ref{eqn:py-solutions}), the set $\phi_{T,p_x}$ can contain zero, one, or a continuum of distributions $P_{x,y}$, depending on the shape and properties of $T$. All of these distributions achieve proportional betting and implement optimal single-letter codes (from Corollary \ref{cor:proportional-betting}). Remarkably, if $\phi_{T,p_x}$ contains more than one element, this implies that {\em optimal performance can be achieved in different channels, even if the strategy is fixed}. Consider the following three cases.

\subsubsection{Overdetermined $p_y$} 
If $|\mathcal{X}| > |\mathcal{Y}|$, the problem is overdetermined. Letting $T^+$ represent the Moore-Penrose pseudoinverse of $T$, a solution exists if $p_x$ is in the row space of $T$, and $p_y = p_x T^+$ is a valid probability; otherwise, $\phi_{T,p_x} = \emptyset$.
This case corresponds to one where there are fewer potential strategies than outcomes -- for example, only two possible portfolios containing three stocks. Physically speaking, it is unlikely that the system would accommodate an arbitrary $p_x$ and the strategy implied by $T$. Thus, in an arbitrary physical situation, it is unlikely that proportional betting is possible, i.e., no single-letter code would be optimal.

\subsubsection{Uniquely determined $p_y$} 
If $|\mathcal{X}| = |\mathcal{Y}|$ (i.e., $T$ is square), then (assuming $T$ is invertible) $p_y = p_x T^{-1}$ is the unique solution if $p_y$ is a valid probability. If $p_y$ is not a valid probability, there are no solutions and $\phi_{T,p_x} = \emptyset$.
In this case, since there are an equal number of potential strategies and outcomes, a solution exists and there is a unique point of equality for the bound in (\ref{eqn:RD-bound}). This is the classical Kelly bet situation, in which the strategy is uniquely matched to the conditional probability $p(x|y)$ and performs proportional betting.

\subsubsection{Underdetermined $p_y$} 
If $|\mathcal{X}| < |\mathcal{Y}|$, the problem is underdetermined. One solution is given by $p_y = p_x T^+$. Further, for any vector $v$ in the null (column) space of $T$, $p_y + v$ is a solution that satisfies $(p_y+v)T = p_x$. (Again, we require that the solutions $p_y + v$ are valid probabilities.)

In the underdetermined case, $P_{x,y}$ is generally different for different vectors $v$ in the null space of $T$, so $\phi_{T,p_x}$ consists of a continuum of joint distributions, characterized by the null (column) space of $T$. Moreover, since the joint distributions are different, they generally have different average distortions $D$. As a result, there is a {\em continuous range over which the bound in (\ref{eqn:RD-bound}) is satisfied with equality.} 

Moving from overdetermined, to uniquely determined, to underdetermined is a matter of adding rows to the strategy matrix $S$. When the problem is underdetermined, proportional betting is achievable over the entire range of distributions in $\phi_{T,p_x}$. This is illustrated by an example in Section \ref{sec:examples}.

\subsection{Properties of $\phi_{T,p_x}$ and the reward matrix $R$}
\label{sec:nonsquare}

The previous section considered how the shape and structure of the strategy matrix affects the results from Proposition \ref{lem:decomposition} and Theorem \ref{lem:RD-inequality}. Here, we can consider the same questions for the reward matrix $R$. The bound in (\ref{eqn:RD-bound}) can be applied, but with a penalty arising from the fact that the decomposition from Proposition \ref{lem:decomposition} is no longer unique. \rev{Moreover, we can also show the conditions under which adding an action (i.e., a phenotype) to $R$ lowers the bound, making better rate-distortion tradeoffs possible.}

Suppose $|\mathcal{Z}| < |\mathcal{X}|$, so that the reward matrix $R$ is nonsquare. Proposition \ref{lem:decomposition} found a decomposition $R = BQ$ where $Q$ is positive and diagonal, and $B$ is row-stochastic; if $R$ is nonsquare, then the setting $Q = \diag(R^+ \vec{1}_{|\mathcal{Z}|})^{-1}$ and $B = R Q^{-1}$ is not the only setting of $Q$ and $B$ with these properties. To see this, let $\mathrm{null}(R)$ represent the null (row) space of $R$, let $v \in \mathrm{null}(R)$ be a vector in that space, let $\hat{Q} = \diag(R^+ \vec{1}_{|\mathcal{Z}|} + v)^{-1}$, and let $\hat{B} = R \hat{Q}^{-1}$. Then we can show $\hat{B}$ is row stochastic for any $v$: 
\begin{align}
    \label{eqn:nonsquare-1}
    \hat{B}\vec{1}_{|\mathcal{X}|} = R \hat{Q}^{-1} \vec{1}_{|\mathcal{X}|} &= R (R^+ \vec{1}_{|\mathcal{Z}|} + v) \\
    &= R R^+ \vec{1}_{|\mathcal{Z}|} + R v \\
    \label{eqn:nonsquare-2}
    &= \vec{1}_{|\mathcal{Z}|} ,
\end{align}
which follows the argument in the proof of the proposition, along with the fact that $Rv = 0$.

For each $v$, the corresponding $\hat{B}$ and $\hat{Q}$ satisfy $R = \hat{B}\hat{Q}$, each leads to a distortion function of the form of (\ref{eqn:GeneralizedDistortionFunction}), and each of these distortion functions satisfies the bound in Theorem \ref{lem:RD-inequality}. However, each $\hat{Q}$ is different, so each one has a different $\Lambda^*$. This implies the following result, which generalizes Theorem \ref{lem:RD-inequality}:
\begin{thm}
    \label{prop:generalization}
    Let $(R,S,p_x)$ represent an investment game, where $R$ is a $|\mathcal{Z}| \times |\mathcal{X}|$ matrix, with $|\mathcal{Z}| < |\mathcal{X}|$. For each vector $v \in \mathrm{null}(R)$, let $\hat{Q}_v = \diag(R^+ \vec{1}_{|\mathcal{Z}|} + v)^{-1}$, with diagonal elements $\hat{q}_{xx,v}$. Further, let $\Lambda_v^* = E[\log \hat{q}_{xx,v}]$. Then:
    \begin{align}
        \label{eqn:nonsquare-cor}
        \rate(D) \geq H(X) - D - \min_{v\in\mathrm{null}(R)} \Lambda_v^* .
    \end{align}
    Furthermore, if $\phi_{T,p_x}$ is nonempty, then $T$ is in the row space of $\hat{B} = R \hat{Q}_{v^\prime}^{-1}$, for $v^\prime = \arg\min_{v\in\mathrm{null}(R)} \Lambda_v^*$.
\end{thm}
\begin{proof}
    To prove (\ref{eqn:nonsquare-cor}), from Theorem \ref{lem:RD-inequality}, (\ref{eqn:RD-bound}) is satisfied for each decomposition that generates $\hat{Q}_v$, so it must also be satisfied for the extreme value in (\ref{eqn:nonsquare-cor}). To prove the statement after the equation, the decomposition in Proposition \ref{lem:decomposition} implies $T = SB$, if it exists. Suppose there exists $v \neq v^\prime$ for which $\phi_{T,p_x}$ is nonempty for $T = S\hat{B} = S R Q_v^{-1}$. Then $\rate(D) = H(X) - D - \Lambda_v^* < H(X) - D - \Lambda_{v^\prime}^*$, which contradicts Theorem \ref{lem:RD-inequality}. Thus, $T$ is in the row space of $\hat{B} = R \hat{Q}_{v^\prime}^{-1}$.
\end{proof}
In Theorem \ref{prop:generalization}, the optimization over $\mathrm{null}(R)$ may be thought of as a penalty for having a reward matrix with rank less than the number of outcomes, or alternatively, {\em a motivation for increasing the rank of $R$:}
\rev{viewing the rows of $R$ as individual phenotypes of the organism, new phenotypes may arise through various mechanisms \cite{kaessmann2010origins},
but previous phenotypes are still available within the population. Thus, an organism with a given set of phenotypes is ``stuck with'' the existing rows in its reward matrix $R$, but {\em it can add new ones}, and these new rows alter the bound in (\ref{eqn:nonsquare-cor}), by pushing it lower and hence allowing the organism to find better-performing strategies. In the following corollary, we show the conditions under which a new row in $R$ can improve the bound in (\ref{eqn:nonsquare-cor}):}

\rev{
\begin{corr}
    \label{corr:AddedRow}
    Let $R$ be a $|\mathcal{Z}| \times |\mathcal{X}|$ matrix, with $|\mathcal{Z}| < |\mathcal{X}|$, and let $\hat{B} = R \hat{Q}_{v^\prime}^{-1}$, for $v^\prime = \arg\min_{v\in\mathrm{null}(R)} \Lambda_v^*$, as in Theorem \ref{prop:generalization}. Further let
    \begin{align}
        R^\prime = \left[ \begin{array}{c} R \\ r \end{array}\right] ,
    \end{align}
    where $r$ is a vector with nonnegative entries. If $R^\prime$ satisfies the conditions of the decomposition in Proposition \ref{lem:decomposition}, and there exist valid strategies $S$ and $T$ such that $S \hat{B} = T$, then
    \begin{align}
        \min_{v\in\mathrm{null}(R)} \Lambda_v^* \leq \min_{v\in\mathrm{null}(R^\prime)} \Lambda_v^* .
    \end{align}
\end{corr}
\begin{proof}
    Suppose the opposite, that $\min_{v\in\mathrm{null}(R)} \Lambda_v^* > \min_{v\in\mathrm{null}(R^\prime)} \Lambda_v^*$.
    If $S \hat{B} = T$, then $T$ is in the row space of $\hat{B}$, and $\phi_{T,p_x}$ is nonempty. Then the bound (\ref{eqn:nonsquare-cor}) is satisfied with equality. Now let
    \begin{align}
        S_0 = \Big[ \begin{array}{c} S \:  \vec{0}  \end{array}\Big] ,
    \end{align}        
    where $\vec{0}$ is an all-zero column vector of the appropriate length; it can then be shown that the $\rate(D)$ curves for $(R,S,p_x)$ and $(R^\prime,S_0,p_x)$ are identical. Thus, there exists a point on the $\rate(D)$ curve for $(R^\prime,S_0,p_x)$ that intersects with the bound for $\min_{v\in\mathrm{null}(R)} \Lambda_v^*$. However, by the initial assumption, the bound for $\min_{v\in\mathrm{null}(R^\prime)} \Lambda_v^*$ lies strictly above the bound for $\min_{v\in\mathrm{null}(R)} \Lambda_v^*$, so there exist points on the $\rate(D)$ curve for $(R^\prime,S_0,p_x)$ that lie below this ``bound''. This contradicts Theorem \ref{prop:generalization}, which proves the corollary.
\end{proof}}

\section{Examples}
\label{sec:examples}

\rev{In this section we present examples. We first give three illustrative examples to demonstrate Theorem \ref{lem:RD-inequality}, Proposition \ref{prop:bijection}, Theorem \ref{prop:generalization}, and their corollaries. In these first examples, parameters are chosen for clarity of illustration, rather than with any particular physical system in mind. We subsequently give two examples of our results in highly simplified biological scenarios, and give {\em Monte Carlo} results to indicate how the information-theoretic performance bounds can be approached with single-letter codes, even at points far from the Kelly betting points given in the set $\phi_{T,p_x}$. These results allow us to conclude that rate-distortion is a useful tool for understanding biological communication, while motivating future analytical and experimental work. \revvv{Finally, we give an example illustrating how a Kelly-like encoding scheme might be implemented using block codes, in contrast to single-letter codes.}}

Code to generate the figures is available online \cite{zenodo}, and this code can be modified to produce similar curves for different parameters. For each example, the $\rate(D)$ curve was produced using the algorithm from \cite{hayashi2023bregman}.

\subsection{Illustrative examples}

\begin{figure}[t!]
    \centering
    \includegraphics[width=0.9\columnwidth]{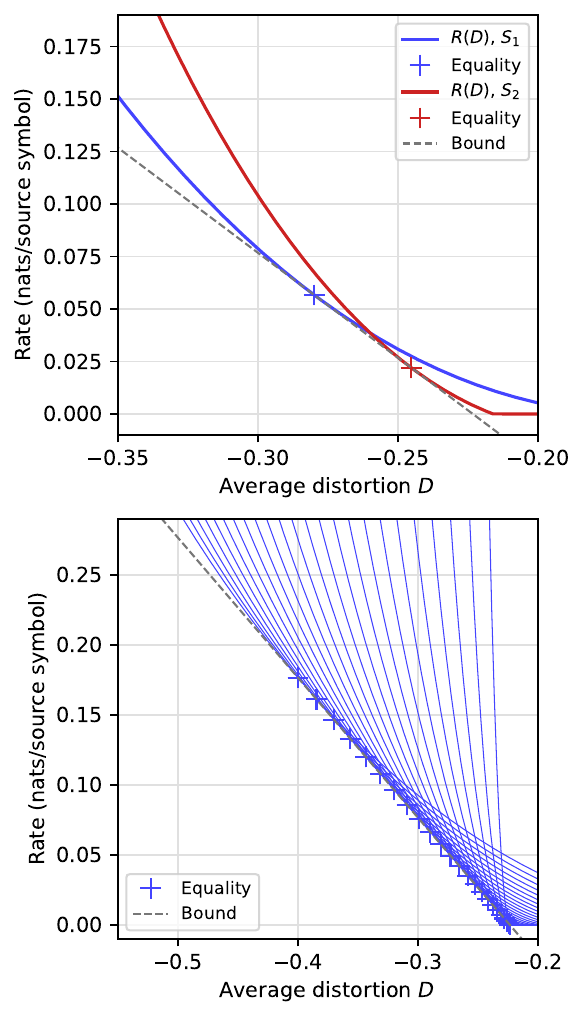}
    \caption{Example illustrating Proposition \ref{lem:decomposition} and Theorem \ref{lem:RD-inequality}. Top subfigure: blue `x' markers represent the investment game from (\ref{eqn:RD-example-R}), while red dots represent the same investment game substituting $S_2$, given in (\ref{eqn:RD-example-newS}), for $S_1$. Points of equality for each curve are represented by large `+' markers. Bottom subfigure: as strategies vary, the $\rate(D)$ curves meet the bound from (\ref{eqn:RD-bound}) at a \rev{sequence} of points, \rev{with points of equality as marked.}}
    \label{fig:Figure2}
\end{figure}

{\em Example 1:} Our first example illustrates Theorem \ref{lem:RD-inequality}, the bound in (\ref{eqn:RD-bound}), and the location where the bound is achieved with equality. Consider an investment game $(R,S_1,p_x)$ with
\begin{align}
    \label{eqn:RD-example-R}
    (R,S_1,p_x) &= 
    \left( \left[
        \begin{array}{cc}
            2 & \frac{1}{2} \\
            \frac{1}{2} & 2
        \end{array}
    \right] , 
    \left[
        \begin{array}{cc}
            \frac{7}{9} & \frac{2}{9} \\
            \frac{2}{9} & \frac{7}{9}
        \end{array}
    \right] , 
    \left[ \frac{1}{2},\: \frac{1}{2} \right] \right).
\end{align}
%
%
The $\rate(D)$ curve (blue line) and bound (\ref{eqn:RD-bound}) (grey line) are illustrated in the top subfigure of Figure \ref{fig:Figure2}, along with the unique point at which $\rate(D)$ meets the bound.
The points of equality may be calculated explicitly from the joint distribution $P_{x,y}$ arising from $P_{x|y} = T$ and $p_y = p_x T^{-1}$, as well as the distortion function (\ref{eqn:GeneralizedDistortionFunction}), where $T = S_1 B^{-1}$ is the effective strategy induced by $S_1$ and the decomposition $R = BQ$. Moreover, it is interesting to note that the convexity of $\rate(D)$ means that the bound can be increasingly loose away from the point of equality.

Continuing this example, it is important to note that both $\rate(D)$ and the point of equality $D^*$ change with the strategy matrix: for example, the red line in the top subfigure of Figure \ref{fig:Figure2} represents $\rate(D)$ with 
\revv{
\begin{align}
    \label{eqn:RD-example-newS}
    S_2 &= \left[
        \begin{array}{cc}
            \frac{4}{5} & \frac{1}{5} \\
            \frac{2}{5} & \frac{3}{5}
        \end{array}
    \right]
\end{align}}
which has a different proportional betting solution than the strategy in (\ref{eqn:RD-example-R}), and hence a different $D^*$. \revv{Unlike $S_1$, it is nonsymmetric}. 

By {\em varying the strategy}, it is possible to achieve equality at different points along the bound, representing each point at which proportional betting is possible. This is further illustrated in the bottom subfigure, which shows the superimposed $\rate(D)$ curves for strategy matrices of the form
\begin{align}
    \label{eqn:VariableStrategyMatrix}
    S &= \left[
        \begin{array}{cc}
            1-a & a \\
            a & 1-a
        \end{array}
    \right]
\end{align}
where $a$ varies in equal increments from $a = 0.02$ to $a = 0.48$. In the figure, the curves each meet the bound at different points, and by superimposing them, they collectively approximate achieving the bound with equality. 

\begin{figure}[t!]
    \centering
    \includegraphics[width=0.9\columnwidth]{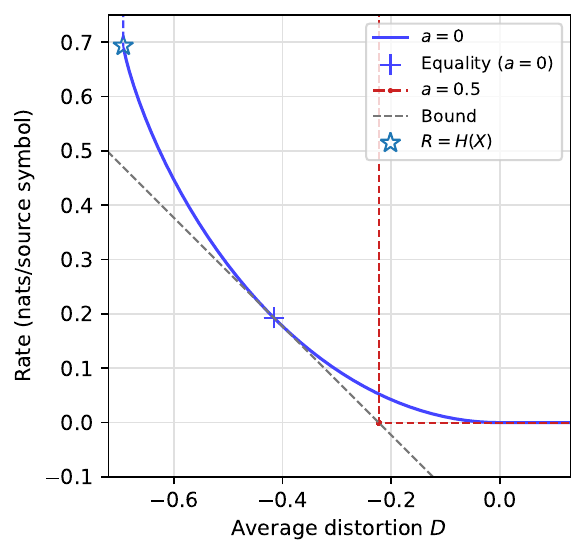}
    \caption{\rev{Illustration of the strategies corresponding to $a = 0$ and $a = 0.5$ in (\ref{eqn:VariableStrategyMatrix}). The bound, the point where the $a=0$ curve meets the bound, and the point where $R = H(X)$ when $a = 0$, are included for comparison.}}
    \label{fig:Figure3}
\end{figure}

\rev{It is also interesting to consider extreme values of $a$ in (\ref{eqn:VariableStrategyMatrix}), which are illustrated in Figure \ref{fig:Figure3}. First consider the case where $a = 0.5$. In this case, both rows of $S$ are the same, so performance is independent of side information. Moreover, this strategy is equivalent to proportional betting with respect to $p_x$, which is the Kelly betting strategy with respect to the prior distribution (i.e., without side information). Thus, the $\rate(D)$ curve corresponding to this point, depicted by a dashed red line, is independent of rate for any rate greater than zero, and the point of equality on this curve is the point at $\rate(D) = 0$.}

\rev{Now consider the case where $a = 0$, i.e., the investor applies their entire fortune to one or the other action. Since the reward matrix $R$ is nondiagonal, from Proposition \ref{lem:decomposition} it can be decomposed as $R = BQ$, where
\begin{align}
    B &= \left[
        \begin{array}{cc}
            0.8 & 0.2 \\
            0.2 & 0.8
        \end{array}
    \right],
\end{align}
and the effective strategy is $T = SB = B$ (since $S$ with $a = 0$ is the identity matrix). The point where the bound is satisfied with equality is indicated in the figure. Remarkably, if the investor is operating at this point, they cannot change their strategy in response to better information, because they are already betting their entire fortune on one action. While the bound therefore cannot be satisfied with equality above and to the left of this point, the single-letter code may still be optimal; for example, the point marked ``$R = H(X)$'' in the figure occurs at the rate where the entire source can be sent losslessly. 
At this point, an optimal approach would be to send the entire source, and use only the best phenotype for each environment; this approach is equivalent to a single-letter code, and appears to lie along the $\rate(D)$ curve for $a = 0$.}

\rev{More generally, as this example demonstrates, proportional betting is {\em sufficient} for a single-letter code to be optimal, it is not {\em necessary}. The required form of the distortion function (\ref{eqn:gastpar_distortion}) admits proportional betting with $c=1$ and $d_0(x) = \Lambda^*(x)$ (as in Corollary \ref{cor:proportional-betting}), but there exist other settings that give optimal single-letter-code solutions. This is further seen through {\em Monte Carlo} simulations presented in the next section, and is a promising avenue of investigation for future work.
}

{\em Example 2:} While the above example illustrated achieving the bound at various points with a {\em variable strategy}, in section \ref{sec:StrategyMatrixProperties} we showed that a similar effect is possible with a {\em fixed strategy}, which we illustrate with our second example. Here we consider the effect of adding additional rows $s^{(y)}$ to the reward matrix $S$, illustrating the overdetermined, uniquely determined, and underdetermined cases. 
In this example, the investment game $(R,S,p_x)$ has (for simplicity) a diagonal reward matrix $R = 3I_3$, where $I_3$ represents the $3 \times 3$ identity matrix; and prior probabilities of the outcomes $p_x = [\frac{1}{3},\frac{1}{3},\frac{1}{3}]$. For $S$, consider the following three strategy matrices, with $|\mathcal{Y}| = 2$, $3$, and $4$, respectively:
\begin{align}
    \nonumber
    S_2 &=
    \left[
        \begin{array}{ccc}
            0.7 & 0.15 & 0.15\\
            0.15 & 0.7 & 0.15
        \end{array}
    \right], \\
    \label{eqn:StrategyMatrices}
    S_3 &=
    \left[
        \begin{array}{ccc}
            & S_2 & \\
            0.15 & 0.15 & 0.7
        \end{array}
    \right], \\
    \nonumber
    S_4 &=
    \left[
        \begin{array}{ccc}
            & S_3 & \\
            0.4 & 0.3 & 0.3
        \end{array}
    \right] ,
\end{align}
noting that $S_2$ is a submatrix of $S_3$, and $S_3$ is a submatrix of $S_4$; thus, $S_3$ and $S_4$ keep the previous strategies and add a new one. 

\begin{figure}[t!]
    \centering
    \includegraphics[width=0.9\columnwidth]{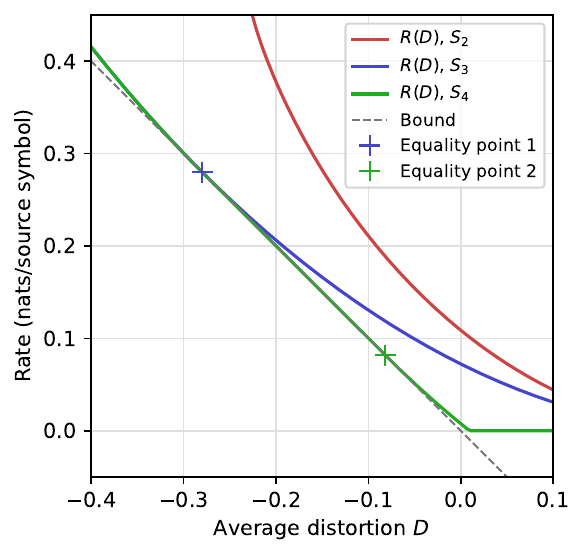}
    \caption{Illustration of the effect of adding strategies. Lines depict $\rate(D)$ for strategy matrices $S_2$, $S_3$, and $S_4$ in (\ref{eqn:StrategyMatrices}), where $S_2$ is a submatrix of $S_3$, and $S_3$ is a submatrix of $S_4$. Using $S_3$, $\rate(D)$ satisfies the bound (\ref{eqn:RD-bound}) with equality at the single point given by the upper \rev{blue} cross \rev{(Equality point 1)}, while using $S_4$, $\rate(D)$ satisfies the bound with equality at {\em all points} between \rev{Equality point 1 and Equality point 2}. Above the upper cross, the lines for $S_3$ and $S_4$ coincide.}
    \label{fig:Figure4}
\end{figure}

Since $R$ is diagonal, $S$ and $T$ are equal: $T_i = S_i$ for $i \in \{2,3,4\}$. It is clear by inspection that $p_x$ is not in the row space of $S_2$ (i.e., $p_x = p_y T_2 = p_y S_2$ is overdetermined); thus, $\phi_{T_2,p_x} = \emptyset$, and there is no solution satisfying proportional betting. For $S_3$, it can be verified that the rows are linearly independent, so there is a unique solution: $p_y = p_x T_3^{-1} = p_y S_3^{-1}$, and $\phi_{T_3,p_x}$ contains a unique distribution corresponding to this solution. For $S_4$, the problem $p_x = p_y T_4 = p_y S_4$ is underdetermined, and $\phi_{T_4,p_x}$ contains a continuum of distributions for which $p_y$ is in the null (column) space of $T_4$. These cases are illustrated in Figure \ref{fig:Figure4}, where the $S_2$ line does not touch the bound from (\ref{eqn:RD-bound}), the $S_3$ line touches it at a single point (the upper cross), and the $S_4$ line touches it at a range of points (all points between the two crosses).
Thus, expanding the null (column) space of $S$ increases the range of $D$ and $\rate(D)$ at which proportional betting is possible.



\begin{figure}[t!]
    \centering
    \includegraphics[width=0.9\columnwidth]{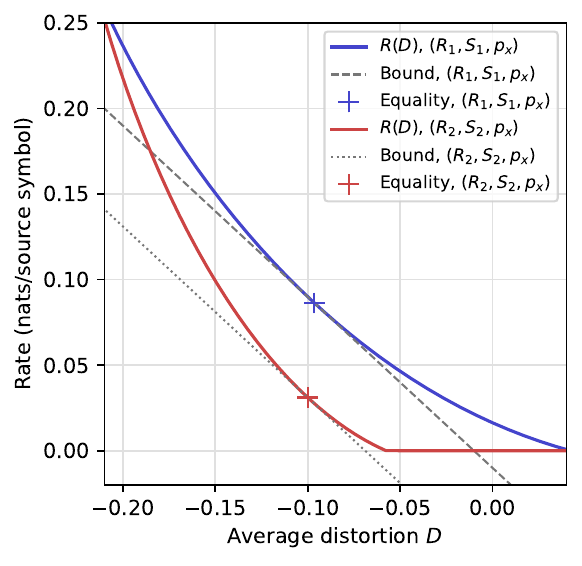}
    \caption{Example illustrating a reward matrix $R_1$ with 2 actions and 3 outcomes, \rev{and the effect of adding an additional row to $R_1$. 
    In the figure, the bound (\ref{eqn:nonsquare-cor}) is given for $R_1$ and $R_2$, while the $\rate(D)$ curves are given for both $(R_1,S_1,p_x)$ in (\ref{eqn:add-row-1}), and $(R_2,S_2,p_x)$ in (\ref{eqn:add-row-2}). Points of equality are indicated.}}
    \label{fig:Figure5}
\end{figure}

{\em Example 3:} The following example illustrates the effect of a nonsquare $R$, \rev{and the effect of adding a new row to $R$}. Let $(R_1,S_1,p_x)$ be an investment game with
\begin{align}
    \nonumber\lefteqn{(R_1,S_1,p_x)}&\\ 
    \label{eqn:add-row-1}
    &=
    \left( \left[ \begin{array}{ccc} 
        2 & 0 & 1 \\
        0 & 2 & 1
    \end{array}\right],
    \left[ \begin{array}{ccc} 
        0.8 & 0.2 \\
        0.2 & 0.8
    \end{array}\right], 
    \Big[ 0.2,\:0.3,\:0.5 \Big] \right).
\end{align}
\rev{Furthermore let $(R_2,S_2,p_x)$ be an investment game with the same $p_x$ as above, and
\begin{align}
    \label{eqn:add-row-2}
    R_2 &= \left[ \begin{array}{ccc} 
        2 & 0 & 1 \\
        0 & 2 & 1 \\
        0.5 & 0.5 & 2
    \end{array}\right] , \:\:\:\:
    S_2 = \left[ \begin{array}{ccc}
        0.4 & 0.1 & 0.5 \\
        0.1 & 0.4 & 0.5
    \end{array}\right]
\end{align}
where $R_2$ is formed by adding the row $r = [0.5,0.5,2]$ to $R$, and the strategy $S_2$ was chosen to satisfy this bound with equality, following Proposition \ref{prop:bijection}. It can be shown that $R_2$ satisfies the conditions of Corollary \ref{corr:AddedRow}, while from Figure \ref{fig:Figure5}, the bound for $R_2$ falls significantly below the bound for $R_1$, and it is clear from the figure that the new, lower bound is achievable with equality.
%
}


\subsection{Biologically-inspired examples}
\label{sec:biologically-inspired-examples}

\rev{Here we present two highly simplified examples inspired by biologically relevant scenarios. These examples illustrate the usefulness of the results and motivate future experiments to validate their predictions.} 

\rev{This section also considers the practicality of single-letter coding, beyond the optimality results we presented above. For example, in Theorem \ref{lem:RD-inequality} and following results, we showed that Kelly betting forms an optimal single-letter code at points in the set $\phi_{T,p_x}$. Now we ask the question: is this bound only achievable rarely, for fine-tuned parameters; or is it common for single-letter codes to approach the bound (either $\rate(D)$ or (\ref{eqn:RD-bound}))? If the latter is true, it means that $\rate(D)$ and (\ref{eqn:RD-bound}) are useful guides to the performance of natural communication systems.}

\rev{To answer this question, we must find the operating point on the $(D,\rate)$ plane corresponding to a given investment game $(R,S,p_x)$ operating in a given noisy channel $P_{y|x}$. The investment game is used to find a distortion function $d(x,y)$ as in (\ref{eqn:GeneralizedDistortionFunction}), while $P_{y|x}$ is used to find the joint distribution $P_{x,y}$. From these quantities, the operating point is given by $(D,\rate) = (I(X;Y),E[d(x,y)])$. We then consider how investment games with given $d(x,y)$ perform with respect to $\rate(D)$ in randomly chosen channels. Remarkably, as our examples show, a given $\rate(D)$ function can be approached closely along its entire length.}


\rev{{\em Example 4:} Consider an environment where the organisms face an {\em infrequent catastrophe}. The organism may express a ``growth'' phenotype, allowing it to grow and thrive under normal conditions, but die once the catastrophe occurs. The organism may also express a ``survival'' phenotype, sacrificing growth under normal conditions but allowing it to survive the catastrophe. This simple model captures some features of antibiotic resistance: a bacteria may enter a dormant state to survive an antibiotic attack; moreover, there are communication elements of this problem, as entering this state may be done in response to a warning signal from another organism \cite{vega2012signaling,niu2024bacterial}.}

\rev{For this example, let
\begin{align}
    R &= \left[ 
        \begin{array}{cl} 
            2 & 0 \\
            1 & 1
        \end{array}
    \right] , 
\end{align}
with the first row representing the ``growth'' phenotype, and the second representing the ``survival'' phenotype; moreover, the first column represents the normal environment, and the second column represents the catastrophic environment. As the catastrophe is assumed to be infrequent, let $p_x = [0.8,0.2]$.}

\rev{Suppose the organism can use one of the following two strategy matrices: 
\begin{align}
    S_1 &= \left[ 
        \begin{array}{cl} 
            0.9 & 0.1 \\
            0.1 & 0.9
        \end{array}
    \right] , \:\:\:\:
    S_2 = \left[ 
        \begin{array}{cl} 
            0.6 & 0.4 \\
            0.4 & 0.6
        \end{array}
    \right] .
\end{align}
Each row of $S$ is selected by the side information; thus, the first and second rows of these matrices reflect side information in which the catastrophe is believed to be {\em less} and {\em more} likely, respectively.
In both $S_1$ and $S_2$, bet hedging occurs which may reflect uncertainty in the side information. However, in $S_1$, a far higher proportion of the population will express the most likely phenotype, either ``growth'' or ``survival''; while in $S_2$, the phenotypes are more balanced, although still slightly biased towards the most likely phenotype.}

\begin{figure}[t!]
    \centering
    \includegraphics[width=0.9\columnwidth]{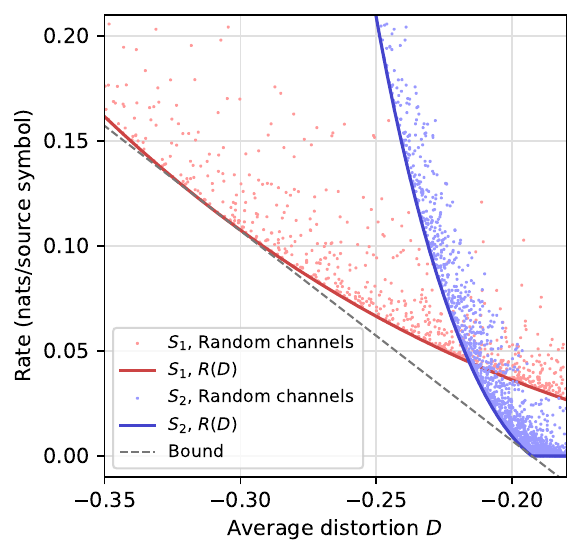}
    \caption{\rev{Example illustrating natural communication system performance in an environment with {\em infrequent catastrophes}. The red curve gives $\rate(D)$ for $S_1$, and the red dots give randomly generated $P_{y|x}$ used with $(R,S_1,p_x)$. The blue curve gives $\rate(D)$ for $S_2$, and the blue dots give randomly generated $P_{y|x}$ used with $(R,S_2,p_X)$.
    Because of the random generation of $P_{y|x}$, approximately half of the randomly-generated channels are cut off to the right of the figure's axes. (These channels give extremely poor performance, for example because they implement a channel where the environment and side information are inverted).
    }}
    \label{fig:Figure6}
\end{figure}

\rev{Results are shown in Figure \ref{fig:Figure6}. The $\rate(D)$ curves are given, along with the bound from (\ref{eqn:RD-bound}). We observe that both $\rate(D)$ curves contact the bound, where $S_1$ contacts the bound at a higher rate than $S_2$. Thus, an organism with high-quality side information would have a higher growth rate with less bet hedging (for example by using $S_1$), while an organism with low-quality side information would do better with more bet hedging (for example by using $S_2$). This is an intuitively correct outcome, but can be quantified by examining the figure.}

\rev{In the figure, we also select random distributions $P_{y|x}$. The distribution $P_{y|x}$ is the communication (or sensing) channel available to the organism as it gathers information about the possible catastrophe. In the same way that a conventional communication channel would vary for a variety of factors (e.g., distance to the transmitter, weather conditions, etc.), the $P_{y|x}$ used in a natural communication scenario will also vary. To illustrate the robustness of a natural communication system, we choose different reliabilities of the channel between the transmitter and receiver, to see how closely a varying system could approach $\rate(D)$. (To reiterate, our results show that the Kelly betting point can achieve exactly $\rate(D)$ at one point, but it is interesting to consider the achievability of $\rate(D)$ elsewhere.) In the figure, for each $x$ we generate $p(y|x) = \frac{u_{x,y}}{\sum_{y} u_{x,y}}$, where $u_{x,y} \in [0,1]$ is a uniformly distributed random variable, independent for each $y$; this method of generating $P_{y|x}$ is used in order to select as large a set of possible distributions as possible. We observe that $\rate(D)$ is closely achievable along its entire length. We predict that an organism would want to get as close to $\rate(D)$ as possible, which suggests that $\rate(D)$, and the associated bound, would be good predictors of the performance of a natural communication system.}

\rev{{\em Example 5:} Consider an environment with {\em two incompatible nutrients}, with different metabolic machinery required for each. An organism may express the genes to metabolize one or the other nutrient; however, they do not express both genes at the same time. A physical example of this may be found in {\em Escherichia coli} ({\em E. coli}) bacteria, which can consume several different sugars as carbon sources, but only one at a time; this process is known as carbon catabolite repression (CCR) \cite{aidelberg2014hierarchy}. CCR requires a decision on the part of the cell as to which metabolic genes to express, which may change with time \cite{mitchell2009adaptive}.}

\rev{To generate a highly simplified model of this system, suppose the sugars are xylose and lactose, and let 
\begin{align}
    R &= \left[ 
        \begin{array}{cc} 
            2 & 0.4 \\
            0.4 & 2
        \end{array}
    \right] ,
\end{align}
with the first and second columns representing a xylose and lactose environment, respectively, while the first and second rows represent expression of xylose and lactose metabolic pathways, respectively.
The diagonal elements indicate population growth through division, in the presence of the correct carbon source; while the off-diagonal elements are nonzero because the genes for each pathway exhibit a small cross-promotion, so the organism has some chance of survival even if it expresses the wrong gene (see \cite[Fig. 2]{aidelberg2014hierarchy}). Also let $p_x = [0.5,0.5]$, i.e., the nutrients are equally likely to be found.}

\rev{We use a simplified collection of strategies compared with the last example. Let
\begin{align}
    S_1 &= \left[ 
        \begin{array}{cc} 
            1 & 0 \\
            0 & 1
        \end{array}
    \right] , \:\:\:\:
    S_2 = \left[ 
        \begin{array}{cc} 
            1 & 0 \\
            0 & 1 \\
            0.5 & 0.5
        \end{array}
    \right] .
\end{align}
That is, $S_1$ is an all-or-nothing bet on either xylose or lactose (i.e. no bet hedging at all); while $S_2$ combines all-or-nothing with an additional bet hedging strategy. Side information, i.e. information about the local concentration of the nutrients, selects which of these strategies the organism uses. There is evidence that {\em E. coli} communicate via quorum sensing to collectively regulate their CCR \cite{ha2018evidence}, so this side information could be transmitted from neighbouring conspecifics.}

\begin{figure}[t!]
    \centering
    \includegraphics[width=0.9\columnwidth]{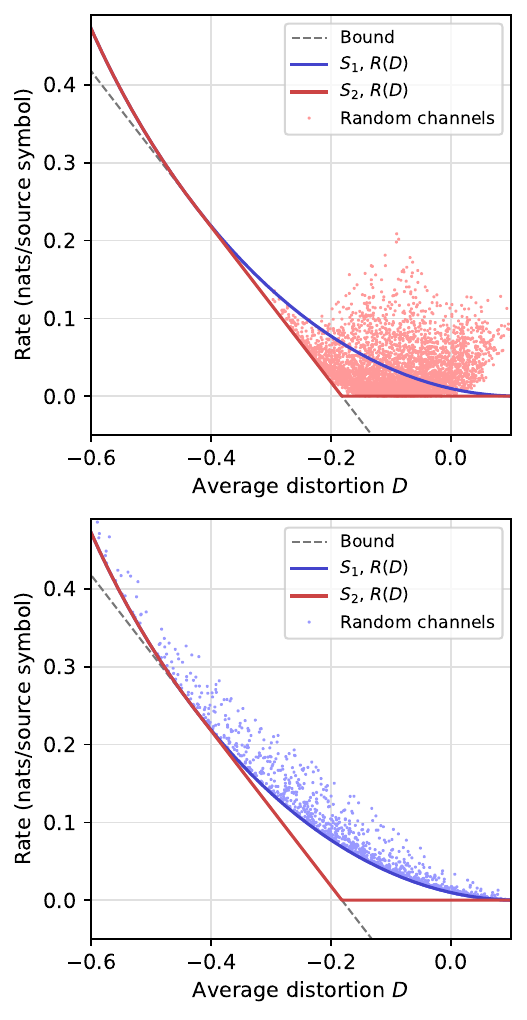}
    \caption{\rev{Example illustrating natural communication system performance in an environment with {\em two incompatible nutrients}. In both subfigures, the red curve gives $\rate(D)$ for $S_1$, while the blue curve gives $\rate(D)$ for $S_2$.
    Using $(R,S_2,p_x)$, red dots in the top subfigure give the operating points for randomly-generated $P_{y|x}$ emphasizing the bet-hedging strategy (see text for details). Blue dots in the bottom subfigure give operating points where the side information leading to the bet-hedging strategy has probability zero, i.e., bet-hedging is never used. As in Figure \ref{fig:Figure6}, approximately half of the randomly-generated channels are cut off to the right of the figure's axes. (These channels give extremely poor performance, for example because they implement a channel where the environment and side information are inverted).
    }}
    \label{fig:Figure7}
\end{figure}

\rev{The results are given in Figure \ref{fig:Figure7}. Consistent with Theorem \ref{lem:RD-inequality} and the discussion following Proposition \ref{prop:bijection}, allowing all-or-nothing bets alongside a bet-hedging strategy (strategy $S_2$) achieves the linear bound for a wide range of rates (red line); moreover, this strategy is always superior to the all-or-nothing strategy $S_1$ (blue line). Moreover, taking the same approach as the above example, we select random $P_{y|x}$ and plot their operating points. As the features of this problem are more complex than the previous example, we illustrate it with two different random distributions of $P_{y|x}$, as follows:
\begin{itemize} 
    \item In the top subfigure, using $S_2$, we use a random side information distribution that emphasizes the bet-hedging strategy. In this figure, for $y \in \{1,2,3\}$ and each $x$, each $P_{y|x}$ is generated as follows: let $p(y=1 | x) = u_x$ and $p(y=2 | x) = v_x$, where $u_x,v_x$ are independent, identically distributed uniform random variables on $[0,0.3]$; and let $p(y=3 | x) = 1-u_x-v_x$. Using these definitions, $E[u_x] = E[v_x] = 0.15$, and $E[1-u_x-v_x] = 0.7$, so this distribution assigns a relatively high probability to $y=3$, the side information that leads to the bet-hedging strategy.
    \item In the bottom subfigure, using $S_2$, we use a side information distribution that excludes the bet-hedging strategy. Here, $P_{y|x}$ is generated in the same was as in Example 4 for $y = 1$ and $y=2$, while $p(y=3|x) = 0$, that is, the side information that leads to the bet-hedging strategy is never observed. Since the side information is never observed, the strategy matrix $S_2$ reduces to $S_1$, and these operating points always lie above both the line for $(R,S_1,p_x)$ and the line for $(R,S_2,p_x)$.
\end{itemize}
We see that the first set closely approaches $\rate(D)$ for $S_2$ at low rate, and the second set closely approaches $\rate(D)$ for both $S_1$ and $S_2$ at high rate. From this, we can predict that an organism would prefer a portfolio of strategies that included {\em both} bet-hedging and all-or-nothing, as that gives the widest possible range of performance over a variety of channels. Moreover, we see that the optimal performance is achievable with single-letter codes for a wide range of rates.
}

\revv{In all of these examples, the reader may wonder how to obtain an upper bound on $\rate(D)$, analogous to the lower bound (\ref{eqn:RD-bound}). In fact, each operating point for a specific $p(y|x)$, for example the dots in Figures \ref{fig:Figure6} and \ref{fig:Figure7}, must lie above $\rate(D)$, as must every point with the same rate and higher distortion, or the same distortion and higher rate; these points can be used to obtain an upper bound. On the other hand, it tends to be more difficult to find lower bounds on $\rate(D)$, another example of which is the Shannon lower bound \cite{shannon1959coding}, which applies to general rate-distortion problems.}

\rev{While these examples are highly simplified ``toy'' examples, they have qualitative biological relevance, and illustrate physical predictions that can be made using the analysis in this paper. 
These results point in promising directions for future experimentation, using mechanisms described in the discussion. Moreover, while these highly simplified examples are biologically plausible, there is no particular limit to the complexity of a real biological system: the range of possible outcomes, the number of actions, and the richness of side information can each be highly complex and represented by large alphabets. Studying larger systems within the framework that we have presented is a promising avenue of future work.}

\subsection{\rev{Example of a Kelly-like block coding scheme}}

Finally, \rev{to illustrate the difference between block and single-letter coding, here we give a tutorial example of a Kelly-like betting scheme implemented using a linear block code, and make the example specific with a Hamming code. Hamming codes are not optimal for this problem, but are chosen for simplicity. The example shows that the computational complexity of any block coding scheme is nontrivial. However, if single-letter codes are optimal, then single-letter coding can obtain {\em the same or better} performance than {\em any} block coding scheme.}


{\em Example 6:} \rev{Consider an investment game $(R,S,p_x)$ with binary outcomes, binary actions, and binary side information. 
The investor has access to a confederate who can communicate side information about the outcomes. The channel between investor and confederate is noiseless but rate-limited: for every $n$ outcomes, the confederate can only communicate $k < n$ bits to the investor. However, the confederate knows {\em in advance} the next $n$ outcomes.}

\rev{Suppose the confederate and investor agree to use a linear block code, such as a Hamming code, with $k \times n$ generator matrix $G$, and $(n-k) \times n$ parity check matrix $H$, as follows:
\begin{enumerate}
    \item The confederate forms a vector of $n$ outcomes, $\vec{x} = [x_1,\ldots,x_n]$.
    \item The confederate uses $\vec{x}$, $H$, and syndrome decoding to find the nearest valid codeword $\vec{y}$. 
    \item The confederate finds a length-$k$ vector $\vec{z}$ satisfying $\vec{z}G = \vec{y}$. Because $\vec{y}$ is a codeword, there must exist such a $\vec{z}$.
    \item The confederate transmits $\vec{z}$ to the investor over the noise-free channel. Since $\vec{z}$ contains $k$ bits to represent $n$ outcomes, this satisfies the confederate's rate constraint.
    \item The investor uses $\vec{y} = \vec{z}G$ as the side information. 
\end{enumerate}
To make the example more concrete, use a Hamming code with $k=4$ and $n=7$, and 
\begin{align}
    H &= 
    \left[ \begin{array}{ccccccc}
        1 & 1 & 0 & 1 & 1 & 0 & 0 \\
        1 & 0 & 1 & 1 & 0 & 1 & 0 \\
        0 & 1 & 1 & 1 & 0 & 0 & 1
    \end{array} \right] ,
\end{align}
where the generator matrix $G$ can be found from $H$ using standard techniques.
Suppose the true vector of outcomes (known to the confederate) is $\vec{x} = [0,1,1,1,1,1,0]$. Using syndrome decoding, we find that $\vec{x}H^T = [1,1,1]$ (recalling that all arithmetic in linear block codes is modulo 2), equal to the fourth column of $H$. Thus, $\vec{y} = [0,1,1,0,1,1,0]$ is the corresponding codeword, and it can be checked that $\vec{z} = [0,1,1,0]$. The confederate subsequently transmits $[0,1,1,0]$ to the investor, and the investor uses $[0,1,1,0,1,1,0]$ as side information. Notice that the side information is incorrect in the fourth position, because that bit was flipped in syndrome decoding. Thus, even though the confederate's knowledge is perfect and the channel is noise-free, the rate constraint and resulting quantization mean that the investor must contend with errors.} 

\rev{We initially assumed that the channel is noiseless. Relaxing that assumption, consider the same problem when the channel is noisy, having capacity $C$: invoking source-channel separation, we could perform the above quantization operation to obtain $\vec{z}$, then encode $\vec{z}$ with a rate-$C$ error-correcting code. If $C = 4/7$ bits per channel use, then the concatenated rate of the system is 1 source letter per channel use, as it is in Kelly betting.} 

\revv{Noncausal access to the source (as in this example) is assumed in the proof of the rate-distortion theorem, in which $\rate(D)$ is shown to be the minimum rate achievable for each distortion (or vice versa). In some scenarios, noncausal access to the source may be possible, such as in the bee-dance example given in the discussion, \revvv{below}. Where noncausal access is not possible, Gastpar's result \cite{gastpar2003code} shows that single-letter codes, which are physically realizable in biological systems, can also achieve $\rate(D)$. This is important because $\rate(D)$ is straightforward to calculate, so it is interesting that it is still relevant to causal and otherwise computationally constrained systems.}

\rev{As part of the publicly available software package to generate the figures in this paper \cite{zenodo}, we include a notebook that allows the user to simulate this process and determine the performance of the given Hamming code scheme.}


\section{Discussion}

\subsection{Optimal single-letter codes are Kelly bets}

While our results showed how Kelly bets are optimal single-letter codes, the converse is also true: that is, for each optimal single-letter code, there exists an investment game for which the system maximizes the expected log growth rate of a multiplicative reward. To be more specific, from (\ref{eqn:DistortionFunction}) and (\ref{eqn:gastpar_distortion}) we need only find $R$ and $S$ so that 
\begin{align}
    \label{eqn:KellyBetTheorem1}
    -\frac{1}{c}E[d(x,y)] = \Lambda^* - H(X|Y) .
\end{align}  
This is satisfied by letting $R$ be a diagonal reward matrix with $r_{xx} = e^{\frac{d_0(x)}{c}}$, and $s_x^{(y)} = p(x|y)$ for all $x$ and $y$. 

From this observation, and the results in \cite{gastpar2003code}, it follows that systems that maximize something other than the expected log growth rate of a multiplicative reward {\em cannot} be implemented with a single-letter code. If such a system were to use a single-letter code, it may be optimized under the constraint of single-letter coding, but additional performance gains could be expected if \rev{block} coding was permitted.

\subsection{Single-Letter Coding versus Sensing}
\label{sec:single-letter-coding-versus-sensing}

The theoretical framework in which these results are derived is one of {\em communication}, in which the sequence of true environments is communicated to the investor across a noisy channel. While many social and natural systems explicitly involve communication, gathering information about the environment can often be viewed as {\em sensing}, as there is no explicit communication strategy employed. Can these results be used to draw conclusions about the efficiency of sensing?

In a sensing task, the environment may be viewed as a trivial ``transmitter'', encoding the true outcome with a ``cue'' \cite{moffett2022minimal,donaldson2010fitness} which can be observed by the sensor. 
Although this setup can be made more complex in various ways, it can also be viewed as a single-letter code with each cue transmitted independently through a noisy channel. 
For example, consider the task of detecting pathogens in the body, which is one task of the immune system. In this example, the true outcome $x$ is the presence or absence of a pathogen.
Pathogens produce antigens which are collected by antigen presenting cells, and can then be recognized by a T cell receptor (TCR), activating an immune response \cite{Irvine2002}. The antigen is a cue: if a pathogen is present, then so are antigens; while if no pathogens are present, no antigens are present either. The process is noisy: the TCR reacts to an antigen only if it encounters one, so depending on antigen and TCR concentration it is possible for the pathogen to be present without activating a TCR; conversely, it is possible (though rare) for the TCR to activate in the absence of an antigen \cite{chang2016initiation,moffett2025comparing}.
The channel output $y$ (TCR activation or nonactivation) is then a noisy view of the current cue (the antigen) representing the current true environment $x$ (presence or absence of pathogen). Many sensing examples similar to this one can be imagined, which are all well described as single-letter codes over noisy channels. If investment games can be formulated for their strategies and rewards, then the bounds and results in Theorems \ref{lem:RD-inequality} and \ref{prop:generalization} can be used to analyze them. Furthermore, since an optimal single letter code is a Kelly bet, natural selection should drive these sensing systems towards optimality.

Finally, recall our assumption that the cost criterion from \cite{gastpar2003code} was always satisfied. The cost function $\rho(x)$ gives the cost of each channel input $x$; in a communication system, the cost may be thought of as ``paid'' by the transmitter, or at least that the transmitter is responsible for picking $p_x$ to optimize the cost. In a sensing system, the sensor is only responsible for guessing $x$, and can't change $p_x$; notionally, the sensor may assume that the transmitter has satisfied the cost criterion, which justifies our assumption in the context of sensing.




\subsection{Experimental confirmation and falsifiability: What might a ``code'' look like in nature?}

Building on observations by Bialek \cite{bialek2012biophysics} and others, we hypothesize that biological communication systems optimize Malthusian fitness, and thus would operate close to their information-theoretic optimum: if not, there would exist a better tradeoff in terms of sensing accuracy and growth rate. \revvv{These observations emphasize the importance of the bound in equation (\ref{eqn:RD-bound}), in that it gives an achievable limit on information processing in natural systems, and that the distance from this bound represents a penalty in terms of inefficiency of information processing. Among other results, we have shown that (\ref{eqn:RD-bound}) is achievable with equality for a wide range of rates: either with changing $S$ (illustrated in the bottom subfigure of Figure \ref{fig:Figure2}), or with fixed but nonsquare $S$ (following Proposition \ref{prop:bijection} and illustrated in Figure \ref{fig:Figure4}). Moreover, as seen in Examples 4 and 5, and illustrated in Figures \ref{fig:Figure6} and \ref{fig:Figure7}, it is feasible for natural communication systems to find better operating points, even if they select channels at random. Thus, we can hypothesize that the bound has {\em predictive power} in describing an organism's sensing apparatus, and that organisms will face selective pressure to move towards the bound \cite{endler1993some,berger2002living}.}

If the hypothesis is correct, then information-theoretic analysis can be used to predict the behaviour of organisms. Experiments can be designed to confirm this hypothesis, for example using chemostats, which have been used in growth rate measurement problems \cite{hoskisson2005continuous}. These devices, which provide a constant nutrient environment while removing members of the population by dilution, maintain bacterial populations at constant growth rates over relatively long periods of time. Recent work has considered the chemostat as an instrument for studying bet-hedging behaviour \cite{wright2020single}, indicating its promise in studying Kelly betting problems. In future work, we believe chemostats would play a significant role in determining the operating points of bacterial populations in changing environments, and in experimentally confirming the information-theoretic results developed in this paper.


\rev{On the other hand, confirmation is not sufficient: a physical hypothesis must be falsifiable, that is, it must be possible to find evidence contradicting the hypothesis. For example, if our hypothesis is false, and the organism is optimizing something other than Malthusian fitness, then Kelly betting and a single-letter code would no longer be optimal (cf. (\ref{eqn:gastpar_distortion})). We have argued that Kelly betting is a useful way to describe biological communication, but it need not be useful in every circumstance: the existence of a code-using biological system in a particular setting, even if the code is simple, would be evidence against our hypothesis in that setting. More generally, the availability of such a test would indicate that our hypothesis is falsifiable.}

\rev{In this direction,} a prime example of a relatively simple inter-organism communication system is the honey bee waggle dance. Through repeated runs of a zig-zagging ``dance,'' a bee with knowledge of the location of a food source  relative to the hive location and the angle of the sun is able to communicate this information to other bees \cite{dyer2002biology}. Each dance run encodes information about the direction of a food source relative to the sun and the distance to the food source in a single run of the dance, with variation between runs. Remarkably, entymologists have provided evidence that bees use a {\em repetition code}, in that bees viewing a waggle dance effectively average over multiple performances to reduce the effects of variation and increase accuracy of location information \cite{tanner2008honey}.  


While we do not analyze the bee waggle dance here, we attempt to demonstrate how our results could be applied to a biological system. Suppose we take a single run of the waggle dance (one zig-zag walk in the direction of the food source followed by a return to the starting position) to be a single ``letter''. In our notation, the true angle and distance to the food source is the true environment $x$, 
and the reconstruction of the angle and distance in the nervous system of the viewing bee is the reconstructed source symbol $y$.
The distortion function $d(x,y)$ relates to the fitness of bees in the hive as a function of the true and reconstructed food source locations. 
If single-letter codes are optimal for this system, then we should be able to reconstruct a corresponding investment game implementing proportional betting. However, if bees implement a repetition code, this might be evidence that single letter codes are far from optimal in this problem. This may be evidence that the distortion function for bee communication is not of the form in (\ref{eqn:gastpar_distortion}).

\section{\revvv{Summary and Conclusions}}

\revvv{While the performance of any communication system is bound by information-theoretic limits, the rate-distortion bound is practical and achievable for naturally-occurring communication systems. The bound can be achieved through Kelly betting, a technique that is closely related to the ``bet-hedging'' behavior that naturally arises from randomness in gene expression; and single-letter codes, which can be implemented with trivial computation. }

\revvv{Our main results address the three open problems given in the introduction. First, we have given a way to concisely determine when Kelly betting is optimal from a source-channel coding perspective, providing an achievable bound on all investment games (Theorem \ref{lem:RD-inequality}, Corollary \ref{cor:proportional-betting}). Second, we have shown that natural information processing can be made robust to environmental changes, by adding rows to the strategy matrix (Proposition \ref{prop:bijection}). Third, we have shown that natural information processing can be made adaptive: under mild conditions, adding new phenotypes improves the achievable bound on investment games (Theorem \ref{prop:generalization}, Corollary \ref{corr:AddedRow}). Because these results quantify fitness gains, an organism subject to evolutionary pressure should achieve them; thus, in nature, we expect to observe biological information-processing systems that make use of these properties. This prediction of biological behavior is an important target for future experimental work.}



\bibliographystyle{ieeetr}
\bibliography{myrefs,bh}

\begin{thebibliography}{10}

\bibitem{gastpar2003code}
M.~Gastpar, B.~Rimoldi, and M.~Vetterli, ``To code, or not to code: Lossy
  source-channel communication revisited,'' {\em IEEE Transactions on
  Information Theory}, vol.~49, no.~5, pp.~1147--1158, 2003.

\bibitem{kelly1956new}
J.~L. Kelly~Jr., ``A new interpretation of information rate,'' {\em Bell System
  Technical Journal}, vol.~35, no.~4, pp.~917--926, 1956.

\bibitem{thorp2008kelly}
E.~O. Thorp, ``The kelly criterion in blackjack sports betting, and the stock
  market,'' in {\em Handbook of asset and liability management}, pp.~385--428,
  Elsevier, 2008.

\bibitem{cover1984algorithm}
T.~Cover, ``An algorithm for maximizing expected log investment return,'' {\em
  IEEE Transactions on Information Theory}, vol.~30, no.~2, pp.~369--373, 1984.

\bibitem{bergstrom2004shannon}
C.~T. Bergstrom and M.~Lachmann, ``Shannon information and biological
  fitness,'' in {\em Information theory workshop}, pp.~50--54, IEEE, 2004.

\bibitem{wu2013interpretations}
B.~Wu, C.~S. Gokhale, M.~van Veelen, L.~Wang, and A.~Traulsen,
  ``Interpretations arising from wrightian and malthusian fitness under strong
  frequency dependent selection,'' {\em Ecology and Evolution}, vol.~3, no.~5,
  pp.~1276--1280, 2013.

\bibitem{raser2005noise}
J.~M. Raser and E.~K. O'shea, ``Noise in gene expression: origins,
  consequences, and control,'' {\em Science}, vol.~309, no.~5743,
  pp.~2010--2013, 2005.

\bibitem{munsky2012using}
B.~Munsky, G.~Neuert, and A.~Van~Oudenaarden, ``Using gene expression noise to
  understand gene regulation,'' {\em Science}, vol.~336, no.~6078,
  pp.~183--187, 2012.

\bibitem{yoshimura1996evolution}
J.~Yoshimura and V.~A. Jansen, ``Evolution and population dynamics in
  stochastic environments,'' {\em Population Ecology}, vol.~38, no.~2,
  pp.~165--182, 1996.

\bibitem{de2023effective}
D.~H. de~Groot, A.~J. Tjalma, F.~J. Bruggeman, and E.~van Nimwegen, ``Effective
  bet-hedging through growth rate dependent stability,'' {\em Proceedings of
  the National Academy of Sciences}, vol.~120, no.~8, p.~e2211091120, 2023.

\bibitem{menu2002bet}
F.~Menu and E.~Desouhant, ``Bet-hedging for variability in life cycle duration:
  bigger and later-emerging chestnut weevils have increased probability of a
  prolonged diapause,'' {\em Oecologia}, vol.~132, pp.~167--174, 2002.

\bibitem{yasui2018bet}
Y.~Yasui and J.~Yoshimura, ``Bet-hedging against male-caused reproductive
  failures may explain ubiquitous cuckoldry in female birds,'' {\em Journal of
  theoretical biology}, vol.~437, pp.~214--221, 2018.

\bibitem{abley2024bet}
K.~Abley, R.~Goswami, and J.~C. Locke, ``Bet-hedging and variability in plant
  development: seed germination and beyond,'' {\em Philosophical Transactions
  of the Royal Society B}, vol.~379, no.~1900, p.~20230048, 2024.

\bibitem{mitchell2009adaptive}
A.~Mitchell, G.~H. Romano, B.~Groisman, A.~Yona, E.~Dekel, M.~Kupiec, O.~Dahan,
  and Y.~Pilpel, ``Adaptive prediction of environmental changes by
  microorganisms,'' {\em Nature}, vol.~460, no.~7252, pp.~220--224, 2009.

\bibitem{miller2001quorum}
M.~B. Miller and B.~L. Bassler, ``Quorum sensing in bacteria,'' {\em Annual
  Reviews in Microbiology}, vol.~55, no.~1, pp.~165--199, 2001.

\bibitem{moffett2022cheater}
A.~S. Moffett, P.~J. Thomas, M.~Hinczewski, and A.~W. Eckford, ``Cheater
  suppression and stochastic clearance through quorum sensing,'' {\em PLOS
  Computational Biology}, vol.~18, no.~7, p.~e1010292, 2022.

\bibitem{striednig2022bacterial}
B.~Striednig and H.~Hilbi, ``Bacterial quorum sensing and phenotypic
  heterogeneity: how the collective shapes the individual,'' {\em Trends in
  Microbiology}, vol.~30, no.~4, pp.~379--389, 2022.

\bibitem{veening2008bet}
J.-W. Veening, E.~J. Stewart, T.~W. Berngruber, F.~Taddei, O.~P. Kuipers, and
  L.~W. Hamoen, ``Bet-hedging and epigenetic inheritance in bacterial cell
  development,'' {\em Proceedings of the National Academy of Sciences},
  vol.~105, no.~11, pp.~4393--4398, 2008.

\bibitem{felsenfeld2014brief}
G.~Felsenfeld, ``A brief history of epigenetics,'' {\em Cold Spring Harbor
  perspectives in biology}, vol.~6, no.~1, p.~a018200, 2014.

\bibitem{veening2008bistability}
J.-W. Veening, W.~K. Smits, and O.~P. Kuipers, ``Bistability, epigenetics, and
  bet-hedging in bacteria,'' {\em Annu. Rev. Microbiol.}, vol.~62, no.~1,
  pp.~193--210, 2008.

\bibitem{gianella2021ecological}
M.~Gianella, K.~J. Bradford, and F.~Guzzon, ``Ecological,(epi) genetic and
  physiological aspects of bet-hedging in angiosperms,'' {\em Plant
  Reproduction}, vol.~34, pp.~21--36, 2021.

\bibitem{donaldson2010fitness}
M.~C. Donaldson-Matasci, C.~T. Bergstrom, and M.~Lachmann, ``The fitness value
  of information,'' {\em Oikos}, vol.~119, no.~2, pp.~219--230, 2010.

\bibitem{donaldson2008phenotypic}
M.~C. Donaldson-Matasci, M.~Lachmann, and C.~T. Bergstrom, ``Phenotypic
  diversity as an adaptation to environmental uncertainty,'' {\em Evolutionary
  Ecology Research}, vol.~10, no.~4, pp.~493--515, 2008.

\bibitem{tal2020adaptive}
O.~Tal and T.~D. Tran, ``Adaptive bet-hedging revisited: Considerations of risk
  and time horizon,'' {\em Bulletin of Mathematical Biology}, vol.~82,
  pp.~1--32, 2020.

\bibitem{berger1971}
T.~Berger, {\em Rate distortion theory: A mathematical basis for data
  compression}.
\newblock Englewood Cliffs, NJ: Prentice-Hall, 1971.

\bibitem{rivoire2011value}
O.~Rivoire and S.~Leibler, ``{The value of information for populations in
  varying environments},'' {\em Journal of Statistical Physics}, vol.~142,
  no.~6, pp.~1124--1166, 2011.

\bibitem{xue2019environment}
B.~Xue, P.~Sartori, and S.~Leibler, ``Environment-to-phenotype mapping and
  adaptation strategies in varying environments,'' {\em Proceedings of the
  National Academy of Sciences U.S.A.}, vol.~116, no.~28, pp.~13847--13855,
  2019.

\bibitem{moffett2022minimal}
A.~S. Moffett and A.~W. Eckford, ``Minimal informational requirements for
  fitness,'' {\em Physical Review E}, vol.~105, no.~1, p.~014403, 2022.

\bibitem{taylor2007information}
S.~F. Taylor, N.~Tishby, and W.~Bialek, ``Information and fitness,'' {\em arXiv
  preprint arXiv:0712.4382}, 2007.

\bibitem{moffett2020fitness}
A.~S. Moffett, N.~Wallbridge, C.~Plummer, and A.~W. Eckford, ``Fitness value of
  information with delayed phenotype switching: Optimal performance with
  imperfect sensing,'' {\em Physical Review E}, vol.~102, no.~5, p.~052403,
  2020.

\bibitem{endler1993some}
J.~A. Endler, ``Some general comments on the evolution and design of animal
  communication systems,'' {\em Philosophical Transactions of the Royal Society
  of London. Series B: Biological Sciences}, vol.~340, no.~1292, pp.~215--225,
  1993.

\bibitem{bialek2012biophysics}
W.~Bialek, {\em Biophysics: searching for principles}.
\newblock Princeton University Press, 2012.

\bibitem{varshney2007optimal}
L.~R. Varshney and D.~B. Chklovskii, ``On optimal information storage in
  synapses,'' in {\em 2007 IEEE Information Theory Workshop}, pp.~408--413,
  IEEE, 2007.

\bibitem{gohari2016information}
A.~Gohari, M.~Mirmohseni, and M.~Nasiri-Kenari, ``Information theory of
  molecular communication: Directions and challenges,'' {\em IEEE Transactions
  on Molecular, Biological and Multi-Scale Communications}, vol.~2, no.~2,
  pp.~120--142, 2016.

\bibitem{tishby2010information}
N.~Tishby and D.~Polani, ``Information theory of decisions and actions,'' in
  {\em Perception-action cycle: Models, architectures, and hardware},
  pp.~601--636, Springer, 2010.

\bibitem{kaessmann2010origins}
H.~Kaessmann, ``Origins, evolution, and phenotypic impact of new genes,'' {\em
  Genome research}, vol.~20, no.~10, pp.~1313--1326, 2010.

\bibitem{golub2013matrix}
G.~H. Golub and C.~F. Van~Loan, {\em Matrix computations}.
\newblock JHU press, 2013.

\bibitem{cover-book}
T.~M. Cover and J.~A. Thomas, {\em Elements of Information Theory 2nd Edition}.
\newblock Wiley-Interscience, July 2006.

\bibitem{marzen2016bio}
S.~E. Marzen, {\em Bio-inspired problems in rate-distortion theory}.
\newblock University of California, Berkeley, 2016.
\newblock Ph.D. dissertation.

\bibitem{verheggen2010alarm}
F.~J. Verheggen, E.~Haubruge, and M.~C. Mescher, ``Alarm pheromones—chemical
  signaling in response to danger,'' {\em Vitamins \& hormones}, vol.~83,
  pp.~215--239, 2010.

\bibitem{martinian2006low}
E.~Martinian and M.~Wainwright, ``Low density codes achieve the rate-distortion
  bound,'' in {\em Data Compression Conference (DCC'06)}, pp.~153--162, IEEE,
  2006.

\bibitem{still2010optimal}
S.~Still, J.~P. Crutchfield, and C.~J. Ellison, ``Optimal causal inference:
  Estimating stored information and approximating causal architecture,'' {\em
  Chaos: An Interdisciplinary Journal of Nonlinear Science}, vol.~20, no.~3,
  p.~037111, 2010.

\bibitem{agrawal2001phenotypic}
A.~A. Agrawal, ``Phenotypic plasticity in the interactions and evolution of
  species,'' {\em Science}, vol.~294, no.~5541, pp.~321--326, 2001.

\bibitem{vembu1995source}
S.~Vembu, S.~Verdu, and Y.~Steinberg, ``The source-channel separation theorem
  revisited,'' {\em IEEE Transactions on Information Theory}, vol.~41, no.~1,
  pp.~44--54, 1995.

\bibitem{berger2002living}
T.~Berger, ``{Living information theory: The 2002 Shannon lecture},'' {\em IEEE
  Information Theory Society Newsletter}, vol.~53, no.~1, pp.~1, 6--19, 2002.

\bibitem{brualdi1966}
R.~A. Brualdi, S.~V. Parter, and H.~Schneider, ``The diagonal equivalence of a
  nonnegative matrix to a stochastic matrix,'' {\em Journal of Mathematical
  Analysis and Applications}, vol.~16, no.~1, pp.~31--50, 1966.

\bibitem{zenodo}
A.~W. Eckford, ``{andreweckford/Distortion-and-fitness: Review release 2},''
  2025.
\newblock https://doi.org/10.5281/zenodo.14845449.

\bibitem{hayashi2023bregman}
M.~Hayashi, ``Bregman divergence based em algorithm and its application to
  classical and quantum rate distortion theory,'' {\em IEEE Transactions on
  Information Theory}, vol.~69, no.~6, pp.~3460--3492, 2023.

\bibitem{vega2012signaling}
N.~M. Vega, K.~R. Allison, A.~S. Khalil, and J.~J. Collins,
  ``Signaling-mediated bacterial persister formation,'' {\em Nature chemical
  biology}, vol.~8, no.~5, pp.~431--433, 2012.

\bibitem{niu2024bacterial}
H.~Niu, J.~Gu, and Y.~Zhang, ``Bacterial persisters: molecular mechanisms and
  therapeutic development,'' {\em Signal transduction and targeted therapy},
  vol.~9, no.~1, p.~174, 2024.

\bibitem{aidelberg2014hierarchy}
G.~Aidelberg, B.~D. Towbin, D.~Rothschild, E.~Dekel, A.~Bren, and U.~Alon,
  ``Hierarchy of non-glucose sugars in escherichia coli,'' {\em BMC systems
  biology}, vol.~8, pp.~1--12, 2014.

\bibitem{ha2018evidence}
J.-H. Ha, P.~Hauk, K.~Cho, Y.~Eo, X.~Ma, K.~Stephens, S.~Cha, M.~Jeong, J.-Y.
  Suh, H.~O. Sintim, {\em et~al.}, ``Evidence of link between quorum sensing
  and sugar metabolism in escherichia coli revealed via cocrystal structures of
  lsrk and hpr,'' {\em Science advances}, vol.~4, no.~6, p.~eaar7063, 2018.

\bibitem{shannon1959coding}
C.~E. Shannon {\em et~al.}, ``Coding theorems for a discrete source with a
  fidelity criterion,'' {\em IRE Nat. Conv. Rec}, vol.~4, no.~142-163, p.~1,
  1959.

\bibitem{Irvine2002}
D.~J. Irvine, M.~A. Purbhoo, M.~Krogsgaard, and M.~M. Davis, ``{Direct
  observation of ligand recognition by T cells.},'' {\em Nature}, vol.~419,
  pp.~845--9, 10 2002.

\bibitem{chang2016initiation}
V.~T. Chang, R.~A. Fernandes, K.~A. Ganzinger, S.~F. Lee, C.~Siebold,
  J.~McColl, P.~J{\"o}nsson, M.~Palayret, K.~Harlos, C.~H. Coles, {\em et~al.},
  ``{Initiation of T cell signaling by CD45 segregation at `close contacts'},''
  {\em Nature Immunology}, vol.~17, no.~5, pp.~574--582, 2016.

\bibitem{moffett2025comparing}
A.~S. Moffett, K.~A. Ganzinger, and A.~W. Eckford, ``{Comparing kinetic
  proofreading and kinetic segregation for T cell receptor activation},'' {\em
  Physical Review Research}, vol.~7, p.~023003, Apr 2025.

\bibitem{hoskisson2005continuous}
P.~A. Hoskisson and G.~Hobbs, ``Continuous culture--making a comeback?,'' {\em
  Microbiology}, vol.~151, no.~10, pp.~3153--3159, 2005.

\bibitem{wright2020single}
N.~R. Wright, N.~P. R{\o}nnest, and N.~Sonnenschein, ``Single-cell technologies
  to understand the mechanisms of cellular adaptation in chemostats,'' {\em
  Frontiers in Bioengineering and Biotechnology}, vol.~8, p.~579841, 2020.

\bibitem{dyer2002biology}
F.~C. Dyer, ``The biology of the dance language,'' {\em Annual Review of
  Entomology}, vol.~47, no.~1, pp.~917--949, 2002.

\bibitem{tanner2008honey}
D.~A. Tanner and P.~K. Visscher, ``Do honey bees average directions in the
  waggle dance to determine a flight direction?,'' {\em Behavioral Ecology and
  Sociobiology}, vol.~62, pp.~1891--1898, 2008.

\end{thebibliography}

\end{document}